\documentclass[journal]{IEEEtran}
\usepackage{color}
\usepackage{hyperref}
\usepackage{graphicx}
\usepackage{amsmath}
\usepackage{graphicx}
\usepackage{times}
\usepackage{epsfig}
\usepackage{url}
\usepackage{caption}
\usepackage{subcaption}
\usepackage{amstext}
\usepackage{amssymb}
\usepackage{amsthm}
\usepackage{color}
\usepackage{amsfonts}
\usepackage{paralist}
\usepackage{algorithm}
\usepackage{algorithmic}
\usepackage{pseudocode}
\usepackage{latexsym}
\usepackage{wasysym}
\usepackage{harmony}
\usepackage{gensymb}
\usepackage{textcomp}
\usepackage{threeparttable}
\usepackage{booktabs}
\usepackage{array}
\usepackage{multirow}
\usepackage{enumitem}
\usepackage{framed}
\usepackage{changes}
\definechangesauthor[name={Vlad}, color=orange]{Vlad}
\definechangesauthor[name={Hanhe}, color=blue]{Hanhe}
\definechangesauthor[name={Tamas}, color=orange]{Tamas}
\definechangesauthor[name={Dietmar}, color=red]{Dietmar}

\ifCLASSINFOpdf
\else
\fi
\begin{document}
%

\title{DeepFL-IQA: Weak Supervision for Deep \\IQA Feature Learning}

%
%

\author{Hanhe Lin, Vlad Hosu, and Dietmar Saupe
	\thanks{H. Lin, V. Hosu and D. Saupe are with the Department of Computer and Information Science, University of Konstanz, 78464 Konstanz, Germany (e-mail: \{hanhe.lin, vlad.hosu, dietmar.saupe\}@uni-konstanz.de).}
}

\markboth{IEEE TRANSACTIONS ON MULTIMEDIA,~Vol.~XX, No.~X, XXX~2020}%
{Shell \MakeLowercase{\textit{et al.}}: Bare Demo of IEEEtran.cls for IEEE Journals}
%



\maketitle
\begin{abstract}

Multi-level deep-features have been driving state-of-the-art methods for aesthetics and image quality assessment (IQA). However, most IQA benchmarks are comprised of artificially distorted images, for which features derived from ImageNet under-perform. 
We propose a new IQA dataset and a weakly supervised feature learning approach to train features more suitable for IQA of artificially distorted images. 
The dataset, KADIS-700k, is far more extensive than similar works, consisting of 140,000 pristine images, 25 distortions types, totaling 700k distorted versions. 
Our weakly supervised feature learning is designed as a multi-task learning type training, using eleven existing full-reference IQA metrics as proxies for differential mean opinion scores. 
We also introduce a benchmark database, KADID-10k, of artificially degraded images, each subjectively annotated by 30 crowd workers. 
We make use of our derived image feature vectors for (no-reference) image quality assessment by training and testing a shallow regression network on this database and five other benchmark IQA databases. 
Our method, termed DeepFL-IQA, performs better than other feature-based no-reference IQA methods and also better than all tested full-reference IQA methods on KADID-10k. 
For the other five benchmark IQA databases, DeepFL-IQA matches the performance of the best existing end-to-end deep learning-based methods on average.

\end{abstract}






%
\IEEEpeerreviewmaketitle

\section{Introduction}


\IEEEPARstart{W}{ith} the rapid advances of electronic technologies, digital images that are captured by users with diverse portable cameras are ubiquitous nowadays. However, these images often contain annoying distortions such as incorrect focus (wrong focal plane), motion blur due to camera shake, or wrong exposure (night time, or high dynamic range in the scene). Therefore, it is particularly important to develop Image Quality Assessment (IQA) models to automatically identify and possibly remove low-quality images from personal collections or online galleries. Moreover, the occurrence of unacceptable distortions may be prevented by quality monitoring and optimal parameter settings during acquisition and encoding for storage. Image enhancement methods may be steered by automated quality assessment to remove artifacts. 

Conventionally, image quality is assessed by groups of human observers. Mean Opinion Scores (MOS) are thus gathered by averaging multiple ratings per stimulus. These subjective studies are usually performed in the lab, are time-consuming, expensive, and cannot be performed in real-time during acquisition or processing. Therefore, subjective image quality assessment has limited applicability in practice, and objective IQA, i.e., algorithmic perceptual quality estimation, has been a long-standing research topic \cite{wang2002image}.

Objective IQA methods are divided into three categories based on the availability of pristine reference images: Full-Reference IQA (FR-IQA) \cite{wang2004image} \cite{zhan2017structural}, Reduced-Reference IQA (RR-IQA) \cite{zhu2019multi}, and No-Reference IQA (NR-IQA) \cite{ye2012unsupervised} \cite{liu2019pre}. Since FR-IQA takes into account information provided by the reference image, it usually outperforms NR-IQA and RR-IQA. Also, NR-IQA is more challenging than FR-IQA and RR-IQA due to the absence of a reference. However, in some applications, reference images are not available and only NR-IQA is feasible.

Recently, with the development of computing technology, deep Convolutional Neural Networks (CNNs) \cite{lecun:2015} are having a profound impact on computer vision research and practice. Instead of carefully designing handcrafted features, deep learning methods automatically discover feature representations from raw image data that are most suitable for a specific task. This has dramatically improved the state-of-the-art in image classification \cite{Simonyan:14c} \cite{he2016deep}, object detection \cite{ren2015faster} \cite{redmon2016you}, and many other domains. 

Although CNNs have achieved remarkable results in many computer vision tasks, developing a CNN-based NR-IQA method is still challenging due to the difficulty of acquiring sufficient data for training. For example, state-of-the-art CNNs like VGG \cite{Simonyan:14c} and ResNet \cite{he2016deep} have many parameters requiring massive amounts of data to train from scratch. However, only recently, the first IQA database with more than 10,000 annotated images for training deep networks have become available (KonIQ-10k, \cite{hosu2019koniq}). Although KonIQ-10k is more than three times as large as the previous largest IQA database, TID2013 \cite{ponomarenko:2015image}, it is still orders of magnitude smaller than ImageNet \cite{deng2009imagenet}, which provided the basis for the immense success of CNNs for many computer vision tasks.

Insufficient labeled data is also a challenge for other fields besides IQA. Weakly supervised learning \cite{zhou2017brief} techniques have been introduced to deal with this in other fields. In this work, we propose a novel weakly supervised approach for NR-IQA. The motivation is that while subjective IQA ratings are difficult to obtain, objective FR-IQA scores are easy to compute, and their performance is generally better than that of NR-IQA methods. Therefore, given a large-scale unrated image set, we train a CNN on corresponding distorted images and their FR-IQA objective scores as labels in order to predict these FR-IQA scores in a no-reference fashion. The trained CNN is expected to perform similar to these FR-IQA methods in applications where reference images are not available. 
Subsequently, we improve image quality prediction by training on a smaller and subjectively annotated IQA database. 

Our primary contributions are the databases we have created for this research and the deep learning techniques we have developed:

\begin{itemize}
\item We introduce a dataset for weakly-supervised learning, consisting of 700,000 images (Konstanz Artificially Distorted Image quality Set, KADIS-700k) without subjective scores. KADIS-700k contains 140,000 pristine images, along with code to generate five degraded versions each by a randomly selected distortion \cite{kadid10k}. In addition, objective quality scores of eleven FR-IQA metrics are computed and provided. KADIS-700k has about 30 times as many pristine images as the largest existing comparable dataset, the Waterloo Exploration Database \cite{ma2017waterloo}. 

\item We introduce another dataset of 10,000 images of a similar kind together with subjective quality scores (Konstanz Artificially Distorted Image quality Database, KADID-10k). KADID-10k consists of 81 pristine images, where each pristine image is degraded by 25 distortions at five levels each. For each distorted image, 30 reliable degradation category ratings are obtained by crowdsourcing, performed by 2,209 crowd workers. Compared to the current largest database of this kind, i.e.,\ TID2013 \cite{ponomarenko:2015image}, KADID-10k is three times as large. Both datasets, together with the source code for the 25 distortions, are available in \cite{kadid10k}.

\item We propose a novel NR-IQA method via weakly supervised deep IQA feature learning, called DeepFL-IQA. Given our large-scale image set with content and distortion diversity, KADIS-700k, eleven FR-IQA metrics are applied to compute objective quality scores for all images. We train a CNN to predict the objective scores of all metrics by our proposed Multi-Task Learning (MTL) framework. 
After training this weakly supervised InceptionResNetV2 model, with the custom MTL head, we extract Multi-Level Spatially Pooled (MLSP) features \cite{Hosu2019Effective} from the locked-in base CNN body. Using these MLSP features, we train another small regression network for subjective score prediction on several benchmark IQA databases\footnote{The source code and pre-trained model will be released in the final version.}. 
\end{itemize}

From a technical standpoint, we introduce the following key improvements:

\begin{itemize}
\item A novel learning approach that joins regression using the HE-norm, a new type of score normalization, inspired by histogram equalization. This improves convergence in the MTL training. 
\item A novel weakly supervised staged training approach, namely DeepFL-IQA. It includes two stages: 1. Fine-tune a CNN, pre-trained on ImageNet, for multi-task score regression, using the HE-norm; 2. Extract MLSP features from the fine-tuned network and train it for NR-IQA using a new regression network.
\end{itemize}

This article is an extension of our KADID-10k database paper \cite{kadid10k_paper}. Distinct contributions lie in that we propose the MLSP-IQA method, evaluate and compare its performance with the state-of-the-art IQA methods on KADID-10k and five other benchmark IQA databases.

\section{Related work}
We divide the related work into two parts: First, we review FR-IQA methods that we applied to generate objective quality scores for weakly supervised learning, then we discuss the current state-of-the-art NR-IQA methods.

\subsection{FR-IQA}

At the first stage of our approach, we need objective quality scores that are generated by FR-IQA methods as ground-truth for the weakly supervised learning. 
Many FR-IQA methods have been proposed in recent years. We collected eleven methods with source code published by their respective authors. 

A typical FR-IQA method works in two steps: First, a similarity map between the reference image and the distorted image is computed, and then the similarity map is converted into a single quality score by a pooling strategy. 

The classical SSIM \cite{wang2004image} uses the luminance, contrast, and structural information to generate a similarity map, from where average pooling is applied to yield the single quality score. 
The performance of SSIM was further improved by MS-SSIM \cite{wang2003multiscale} and IW-SSIM \cite{wang2011information}. The former estimates SSIM on multi-scale resolutions of a given image, and the latter pools the local similarity map by weighting local information content. 
Following a similar idea, several FR-IQA methods that rely on different types of features were proposed. FSIM \cite{Zhang:2011} uses Phase Congruency (PC) and gradient magnitude features. The quality score is estimated by using the PC as a weighting function. Apart from gradient magnitude features, SR-SIM \cite{zhang2012sr} and VSI \cite{zhang2014vsi} use visual saliency as both a feature when computing a local quality map and a weighting function to reflect the importance of a local region when pooling the quality score. 
The SFF index \cite{chang2013sparse} is estimated by combining sparse feature similarity and luminance correlation. 
GMSD \cite{xue2014gradient} uses only the gradient magnitude features, but with a novel standard deviation pooling. SCQI \cite{bae2016novel} characterizes both local and global perceptual visual quality for various image characteristics with different types of structural-distortion.
ADD-GSIM \cite{gu2016analysis} infers an overall quality score by analyzing the distortion distribution affected by image content and distortion, including distributions of distortion position, distortion intensity, frequency changes, and histogram changes. MDSI \cite{nafchi2016mean} generates a gradient similarity map to measure local structural distortion and a chromaticity similarity map to measure color distortion; the final quality score is derived by a novel deviation pooling.

\subsection{NR-IQA}

In terms of the methodology used, NR-IQA can be divided into conventional and deep-learning-based methods. 

\subsubsection{Conventional methods}

Conventional NR-IQA requires domain experts to carefully design a feature extractor that transform raw image data into a representative feature vector from which a quality prediction function is trained to map feature vectors to subjective scores. A number of conventional IQA methods have been proposed in recent years, such as \cite{li2016blind} \cite{liu2018blind}.
Although these methods are not the mainstream of NR-IQA research anymore, they provide a baseline for comparison of NR-IQA methods. 

Many NR-IQA methods are based on statistical properties of images, deriving from the Natural Scene Statistics (NSS) model \cite{srivastava2003advances}. BIQI \cite{Moorthy:2010}, DIIVINE \cite{Moorthy:2011}, and SSEQ \cite{Liu:2014b} have a two-step framework: image distortion classification followed by distortion specific quality assessment. 
BRISQUE \cite{bris} uses scene statistics of locally normalized luminance coefficients to quantify the loss of naturalness in the image due to the presence of distortions. BLIINDS-II \cite{blind2} extracts features based on an NSS model of the discrete cosine transform of the image.

Another approach is the Bag-of-Words (BoW) model using local features. The main idea is to learn a dictionary by clustering raw image patches extracted from a set of unlabelled images, and an image is represented with a histogram by softly assigning raw image patches to the dictionary by a pooling strategy. 
CORNIA \cite{ye2012unsupervised} constructs a codebook with a fixed size of 10,000 using local patches from the CSIQ database \cite{larson:2010most}. To represent an image, it extracts 10,000 7-by-7 patches from it and performs a soft-assignment for each patch to the codebook blocks, using two different metrics. Each patch is thus characterized by two vectors of 10,000 assignment values. The image then is represented by the corresponding max-pooled vectors. Similarly, HOSA \cite{hosa} constructs a smaller codebook of size 100 also for the CSIQ databases by K-means clustering of normalized image patches. Patches are compared to the cluster centroids and softly assigned to several nearest clusters. Aggregated differences of high order statistics (mean, variance, and the skew present) are combined to result in a quality-aware image representation. In comparison to CORNIA, HOSA generally improves performance while reducing the computational cost. 



\subsubsection{Deep-learning-based methods}

Instead of designing handcrafted features, deep-learning-based NR-IQA methods strive to automatically discover representations from raw image data that are most suitable for predicting quality scores.

Current IQA databases appear too small to train deep learning models from scratch. To address this limitation, data augmentation has been applied. For example, in \cite{conv1}, \cite{yan2019naturalness}, and \cite{bosse2018deep}  a CNN was trained on sampled 32-by-32 RGB patches, that were assumed to have the same quality score as the entire image. The quality of an image was estimated by averaging the predicted quality scores of sampled image patches. This approach allows training a CNN from scratch. However, it ignores the semantics of the content depicted in the image, although IQA is content dependent \cite{siahaan2016does}.

Deep transfer learning \cite{DBLP:journals/corr/abs-1808-01974} has also been applied to overcome the lack of training data. It aims to improve the performance of IQA by transferring latent knowledge learned from other image processing tasks that have sufficient data to train, e.g., object classification. DeepBIQ \cite{bianco2016use} extracts deep features of the last fully-connected layer from a fine-tuned Caffe network architecture, pre-trained on ImageNet \cite{deng2009imagenet} for object classification. Their best method predicted image quality by pooling the predictions relying on features of several 227-by-227 patches. This is an early attempt, where the network inputs are image regions of sizes in between the small 32-by-32 pixels patches used in \cite{conv1,bosse2018deep} and full images. 
The authors of \cite{talebi_nima_2017} proposed the NIMA method, which further improved performance, having considered largely the entire image when applying traditional transfer learning using modern CNNs. 
They also showed that the design of the base network architecture is essential for performance.

Given a VGG16 network \cite{Simonyan:14c} with pre-trained weights on ImageNet \cite{deng2009imagenet}, the BLINDER \cite{gao2018blind} model extracts features from the output of each layer. The layer-wise features are fed into SVR models trained to predict the quality score of the input images. The overall image quality is obtained by averaging the predicted quality scores. The approach works well on different IQA databases, even when compared to direct fine-tuning, and is more efficient: extracting features from an image requires a single forward pass through the network, and training the SVR model is faster than fine-tuning the network. Nonetheless, the method assumes little domain shift between the source domain (ImageNet) and the target domain (IQA database). This assumption does not hold in our case, where artificially distorted images have very different types of distortions compared to the ones present in ImageNet. The latter is mostly made up of relatively high-quality images that are meant to be used for object classification.

The method RankIQA \cite{Liu_2017_ICCV} trains a Siamese Network to rank the quality of image pairs, which are easily generated via artificial distortions. The quality score of a single image is predicted by fine-tuning a CNN starting with the weights of the base model used as part of the Siamese Network. It was shown that using more information about image degradation (pair-wise ranking) improves performance. This suggests that other types of relative information, such as the magnitude of the distortion between the reference and distorted image, could further improve performance. This is something we have considered in our work, framing this information transfer as a type of weakly supervised learning.

The MEON model was proposed in \cite{ma2018end}. It contains two sub-networks, a distortion identification network and a quality prediction network. Both of these networks share weights in the early layers. Since it is easy to generate training data synthetically, it first trains the distortion identification sub-network to obtain pre-trained weights in the shared layers. With the pre-trained early layers and the outputs of the first sub-network, the quality score of a single image is estimated after fine-tuning on IQA databases. Similar to RankIQA, the method in \cite{ma2018end} uses partial distortion information during the weakly supervised learning stage, about the type of degradation.

The BPSOM model \cite{pan2018blind} trains three separate generative networks to predict quality maps of images instead of single quality scores. The ground truth quality maps are estimated by three FR-IQA metrics, i.e.,\ SSIM \cite{wang2004image}, FSIM \cite{Zhang:2011}, and MDSI \cite{nafchi2016mean}. 
Predicted maps are regressed into a single quality score by a deep pooling network. By using more information about the distortion magnitude and distribution (quality maps), the BPSOM method achieves a further performance improvement compared to RankIQA. In contrast to BPSOM, we rely on magnitude information obtained from eleven FR-IQA methods, hypothesizing that each method has some advantages over the others, and the aggregate information would be more useful for weakly supervised learning.

Adversarial learning is introduced in \cite{lin2018hallucinated}.  It trains a quality-aware generative network that estimates the reference image for a given distorted image, and an IQA-discriminative network to distinguish such fake reference images from the real reference images. By jointly training both networks, the quality-aware generative network would generate fake reference images that are not able to be recognized by the IQA-discriminate network. 
For quality assessment, given a distorted image and its ``hallucinated'' reference image, a discrepancy map is computed. The discrepancy map, together with high-level semantic information from the generative network, is used to regress a quality score. 

Although deep-learning-based NR-IQA methods have been implemented successfully and achieve better performance than conventional IQA methods, they do not yet generalize well to other databases that they were not trained on. This is due to the relatively small training databases developed until now. In order to fully benefit from our extensive database KADID-700k of artificially distorted images, we need to combine the strengths of existing methods, to more reliably use information about distortion magnitude, and to train more efficiently:

\begin{itemize}
    \item Similar to the BPSOM method, we use information about the distortion magnitude for each artificially distorted image. While BPSOM uses spatial quality maps, we use aggregated global scores; this allows us to train faster on our larger dataset, and facilitates multi-task learning, by combining the information produced by a set of FR-IQA methods.
    \item We use MLSP features that have been adapted to the correct domain by first fine-tuning modern ImageNet architectures on our large KADID-700k dataset, making the training of the second stage very efficient.
    \item We introduce a new type of normalization that allows us to better train from multiple sources of information (multiple FR-IQA scores per image): histogram equalization.
\end{itemize}

\section{KADID-10k and KADIS-700k}

\begin{figure}[t]
\centering
\includegraphics[width=0.48\textwidth]{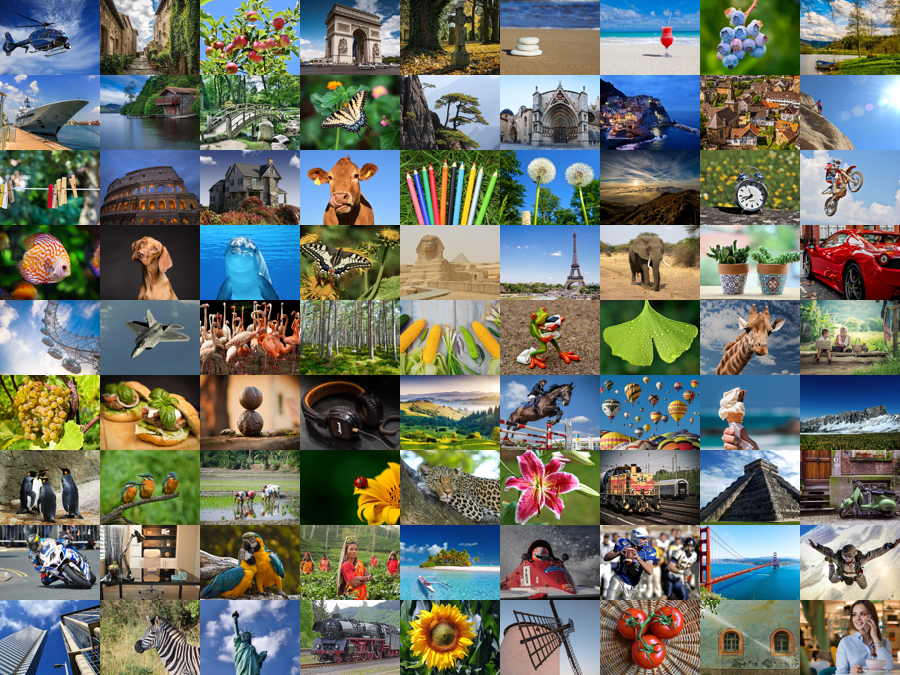}
\caption{The 81 pristine reference images in KADID-10k.}
\label{fig:pristine_81}
\vspace{-10pt}
\end{figure}


To facilitate the development of deep-learning-based IQA methods, we have created two datasets, the Konstanz Artificially Distorted Image quality Database (KADID-10k) and the Konstanz Artificially Distorted Image quality Set (KADIS-700k). KADID-10k contains 81 pristine images and $81 \cdot 25 \cdot 5 = 10125$ distorted images with 30 crowdsourced subjective quality scores each. KADIS-700k contains 140,000 pristine images and 700,000 distorted images. 

\subsection{Dataset creation}

We collected the pristine images for both datasets from Pixabay.com, a website for sharing photos and videos. These images were released under the Pixabay License and were free to be edited and redistributed. Moreover, each uploaded image had been presented to Pixabay users, who cast their votes for accepting or declining it according to its perceptual quality. For each image, up to twenty independent votes had been collected to make a decision. Therefore, the quality rating process carried out by Pixabay provides a reasonable indication that the released images are pristine.

We downloaded 654,706 images with resolutions higher than 1500$\times$1200. All the images were rescaled to span (one dimension fits, the other is larger or equal) the same resolution as that in TID2013 (512$\times$384), maintaining the pixel aspect ratios, followed by cropping to exactly fit the final resolution, if required. With the meta-information we acquired, we arranged these images in descending order of $\log(\text{No. of favorites})/\log(\text{No. of views})$~\cite{schwarz2018will}, from which the top 100 images were selected as candidates of pristine images in KADID-10k. 
After removing 19 images that appeared heavily edited or had similar content, we obtained 81 high-quality pristine images as reference images in KADID-10k, see Fig.~\ref{fig:pristine_81}. As can be seen, the reference images are diverse in content.  
From the remaining images, we randomly sampled 140,000 images as pristine reference images for KADIS-700k.

\begin{figure}[htb]
\centering
\includegraphics[width=0.48\textwidth]{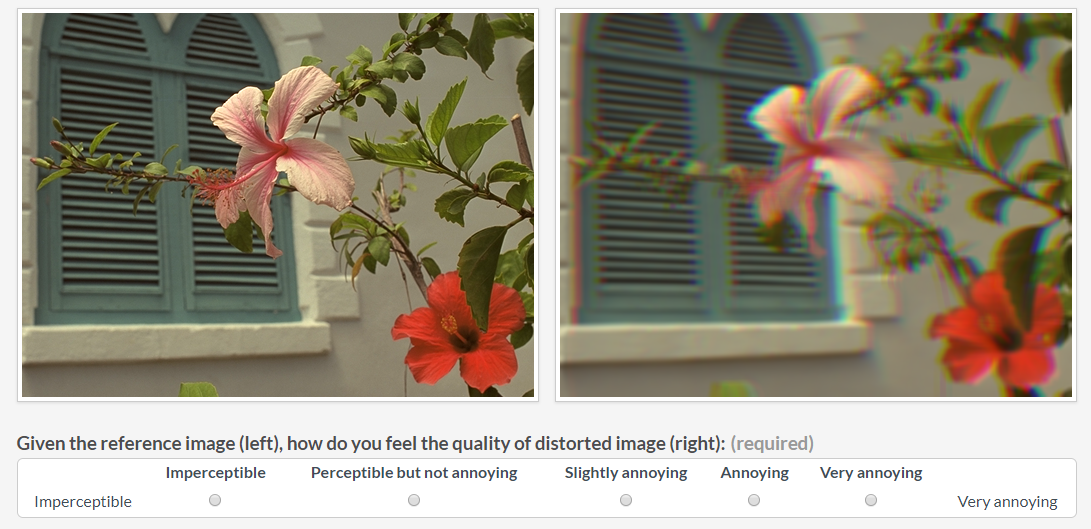}
\caption{User interface for the subjective IQA study.}
\label{fig:kadid10k_interface}
\vspace{-5pt}
\end{figure}

For both of our datasets, we applied 25 image distortion methods. We used two major types of distortions: 1. The first set of 13 distortion types are taken from those used in TID2013. These are indicated by \textit{bold font ids} in the list below; 2. A set of 12 new types are meant to simulate authentic distortions that are frequently encountered in-the-wild. For instance, we propose to use lens and motion blur in addition to Gaussian blur from TID2013. The former distortions are more likely to occur in the wild due to de-focus and camera shake, whereas Gaussian blur is mostly a result of image editing. 

In total there are 25 types of distortions, which can be grouped into:
\begin{itemize}
    \item Blurs
    \begin{description}
        \item \textbf{\#01} Gaussian blur: filter with a variable Gaussian kernel.
        \item \#02 Lens blur: filter with a circular kernel.
        \item \#03 Motion blur: filter with a line kernel.
    \end{description}
    \item Color distortions
     \begin{description}
       \item  \#04 Color diffusion: Gaussian blur the color channels (a and b) in the Lab color-space.
       \item \#05 Color shift: randomly translate the green channel, and blend it into the original image masked by a gray level map: the normalized gradient magnitude of the original image.
       \item \#06 Color quantization: convert to indexed image using minimum variance quantization and dithering with 8 to 64 colors.
       \item \textbf{\#07} Color saturation 1: multiply the saturation channel in the HSV color-space by a factor.
       \item \#08 Color saturation 2: multiply the color channels in the Lab color-space by a factor.
    \end{description}
    \item Compression
    \begin{description}
       \item \textbf{\#09} JPEG2000: standard compression.
       \item \textbf{\#10} JPEG: standard compression.
    \end{description}
    \item Noise
    \begin{description}
        \item \textbf{\#11} White noise: add Gaussian white noise to the RGB image.
        \item \textbf{\#12} White noise in color component: add Gaussian white noise to the YCbCr converted image (both to the luminance ‘Y‘ and the color channels ‘Cb‘ and ‘Cr‘).
        \item \textbf{\#13} Impulse noise: add salt and pepper noise to the RGB image.
        \item \textbf{\#14} Multiplicative noise: add speckle noise to the RGB image.
        \item \#15 Denoise: add Gaussian white noise to RGB image, and then apply a denoising DnCNN to each channel separately.
    \end{description}
    \item Brightness change
    \begin{description}
       \item \#16 Brighten: non-linearly adjust the luminance channel keeping extreme values fixed, and increasing others.
       \item \#17 Darken: similar to brighten, but decrease other values.
       \item \textbf{\#18} Mean shift: add constant to all values in image, and truncate to original value range.
    \end{description}
    \item Spatial distortions
    \begin{description}
       \item \#19 Jitter: randomly scatter image data by warping each pixel with random small offsets (bicubic interpolation).
       \item \textbf{\#20} Non-eccentricity patch: randomly offset small patches in the image to nearby locations.
       \item \#21 Pixelate: downsize the image and upsize it back to the original size using nearest-neighbor interpolation in each case.
       \item \textbf{\#22} Quantization: quantize image values using N thresholds obtained using Otsu's method.
       \item \textbf{\#23} Color block: insert homogeneous random colored blocks at random locations in the image.
    \end{description}
    \item Sharpness and contrast
    \begin{description}
       \item \#24 High sharpen: over-sharpen image using unsharp masking.
       \item \textbf{\#25} Contrast change: non-linearly change RGB values using a Sigmoid-type adjustment curve.
    \end{description}
\end{itemize}
For more information regarding distortions and their source code, please refer to \cite{kadid10k}.

We manually set the parameter values that control the distortion amount such that the perceptual visual quality of the distorted images varies linearly with the distortions parameter, from an expected rating of 1 (bad) to 5 (excellent). The distortion parameter values were chosen based on a small set of images and then applied to all images in both datasets.

\subsection{Crowdsourcing subjective IQA}
Conducting a subjective lab study on 10,125 images is time-consuming and costly; instead, we performed the study on figure-eight.com, a crowdsourcing platform. The experiment first presented workers with a set of instructions, including the procedure to rate an image and examples of different ratings. 
Due to the weakly controlled nature of the experimental setup, including factors such as variable screen resolutions and screen contrast, ambient illumination, etc., crowdsourced subjective studies are less reliable than lab-based ones. We used a standard degradation category rating (DCR) method \cite{ITU:2009} to reduce these effects.

The user interface for our subjective IQA study is illustrated in Fig.~\ref{fig:kadid10k_interface}.
Specifically, given the pristine image on the left side, the crowd workers were asked to rate the distorted image on the right side in relation to the pristine reference image on a 5-point scale, i.e., imperceptible (5), perceptible but not annoying (4), slightly annoying (3), annoying (2), and very annoying (1). To control the quality of crowd workers, we annotated several ``annoying" images and images with ``imperceptible" degradation as test questions according to distortion parameter settings. The test questions served two purposes. First, they filtered out unqualified crowd workers. Before starting the actual experiment, workers took a quiz consisting entirely of test questions. Only those with an accuracy of over 70\% were eligible to continue. Second, hidden test questions were presented throughout the experiments to encourage workers to pay full attention.

\begin{figure}[!htbp]
\centering
\includegraphics[width=0.90\linewidth]{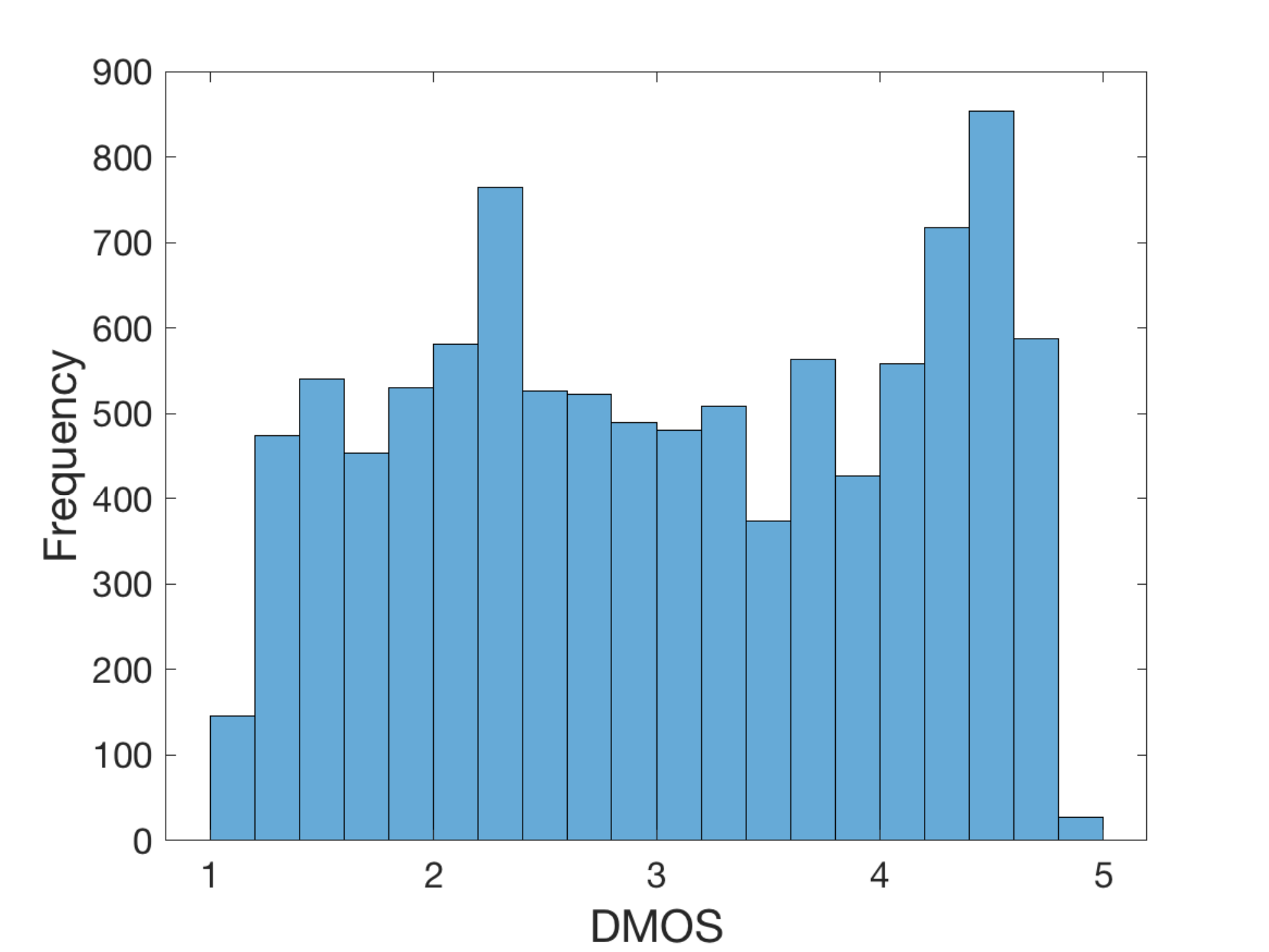}
\caption{Histogram of DMOS in KADID-10k database.}
\label{fig:dmos_hist}
\vspace{-10pt}
\end{figure}

30 DCRs were collected for each image, performed by 2,209 crowd workers. 
This resulted in a DMOS for each image by averaging the DCRs. 
Fig.~\ref{fig:dmos_hist} displays the DMOS histogram of KADID-10k.


\subsection{Reliability analysis}

We had included in our crowdsourcing experiments a few images from the TID2013 IQA database \cite{ponomarenko:2015image}. Thus, we can compare our subjective IQA results with those obtained in the original experiments on TID2013.
In detail, 8 pristine images, along with their distorted images (24 distortions, 5 levels per image) contained in TID2013, were chosen to conduct the same subjective study as for KADID-10k. The SROCC between our crowdsourcing study and the original study is 0.923. This is similar to what was reported in \cite{ponomarenko:2015image} for IQA, where the SROCC between the lab study and an internet study was 0.934. 

We also evaluated the quality of our experimental results by calculating standard reliability measures, the Intra-class Correlation Coefficient (ICC), inter-group correlations, and corresponding error measures. 

In \cite{hosu2018expertise}, reliability measures for IQA experiments using absolute category ratings (single stimulus) were reported. It was shown that the ICC ranges from 0.4 in crowdsourcing experiments, to 0.58 for domain experts (online experiments), and 0.68 in the lab on the CID2013 database \cite{virtanen2015cid2013}. The ICC computed on KADID-10k is 0.66, at the higher end of the reported reliability range. However, our experiments are paired comparisons, which are generally considered easier to perform, and might result in a higher ICC.

With respect to inter-group correlations, computed by bootstrapping with resampling (100 times) from 30 ratings per pair collected in KADID-10k, similar to \cite{hosu2018expertise}, we obtained a very high agreement of 0.982 SROCC and low mean differences of 0.148 MAE and 0.193 RMSE between groups.

\subsection{Database summary}
\begin{framed}
\begin{description}[align=left]
\item [KADID-10k] 81 pristine images, 10,125 distorted images with 30 degradation category ratings (DCR) each. 
\item [KADIS-700k] 140,000 pristine images, 700,000 distorted images with eleven FR-IQA scores each.
\end{description}
\end{framed}

\begin{figure*}[!htbp]
\centering
\includegraphics[width=0.7\textwidth]{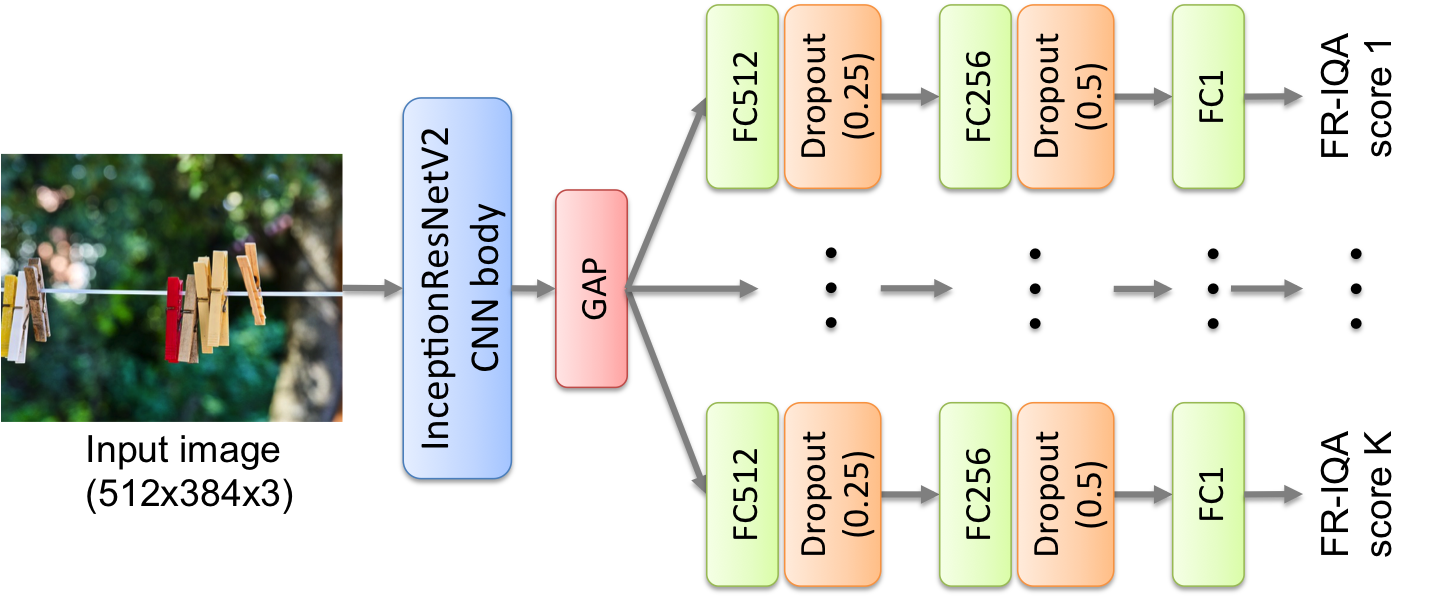}
\caption{
The architecture for multiple FR-IQA score prediction. 1,536 feature maps are produced by passing the input image through the InceptionResNetV2 CNN body. The global average pooling (GAP) layer pools each feature map. The resulting 1,536-dimensional feature vector passes through $K$ separate modules, predicting the $K$ FR-IQA scores. Each module consists of five layers: fully-connected (FC) with 512 units, dropout with rate 0.25, FC with 256 units, dropout with rate 0.5, and output with one neuron.
}
\label{fig:mtl-arch}
\vspace{-5pt}
\end{figure*}

\section{Weakly supervised deep IQA feature learning}

\subsection{Overview}
Our approach consists of two stages. In the first stage, Multi-Task Learning (MTL) is applied to predict objective quality scores of multiple FR-IQA metrics simultaneously. 
Histogram equalization is introduced as used to improve learning. 
The MTL framework allows training a deep model on a large-scale unrated image set, e.g., KADIS-700k. 
In the second stage, Multi-Level Spatially Pooled (MLSP) \cite{Hosu2019Effective} features are extracted from the fine-tuned CNN model, from the first stage, to represent a given image. For the given IQA databases, e.g., KADID-10k, a new regression network is trained to non-linearly map the MLSP features to their corresponding subjective quality scores.

\subsection{Multi-task learning for FR-IQA score prediction}

Assume we have created a large image set $\{(I_n^{r},I_n^{d})~|~n=1,\ldots, N\}$ of $N$ image pairs, where $I_n^r$ and $I_n^d$ correspond to $n$-th reference image and its distorted image, respectively. 
Let $q_n^k = \text{M}_{k}(I_n^r,I_n^d)$ be the objective quality score of the $k$-th FR-IQA method for the $n$-th image pair, where $k=1,\ldots,K$. Our objective is to learn a function $f_\theta(\cdot)$, parameterized by $\theta$, such that
\begin{equation}
\begin{split}
f_\theta(I_n^{d}) \approx (q_n^1,\ldots,q_n^K),~~ n = 1,\ldots, N.
\end{split}
\end{equation}

Since the objective quality scores of FR-IQA methods are highly correlated with subjective quality scores, the learned function is expected to be highly correlated with the subjective quality scores as well. To this end, we perform multi-task learning with deep CNNs to model the described function.

\subsubsection{The proposed architecture}

 The architecture of our proposed deep CNN model for MTL is illustrated in Fig.~\ref{fig:mtl-arch}. We feed an image into the CNN base of InceptionResNetV2 \cite{szegedy2017inception} and use Global Average Pooling (GAP) for each feature map. The resulting feature vector passes through $K$ separate task modules, predicting the $K$ FR-IQA scores. Each task module consists of five layers. These are fully-connected (FC) with 512 units, dropout with rate 0.25, FC with 256 units, dropout with rate 0.5, and output with one neuron.

In our work, the deep CNN model aims to predict the objective scores from $K=11$ available FR-IQA methods.
These methods are SSIM \cite{wang2004image}, MS-SSIM \cite{wang2003multiscale}, IW-SSIM \cite{wang2011information}, MDSI \cite{nafchi2016mean}, VSI \cite{zhang2014vsi}, FSIM$_{\text{c}}$ \cite{Zhang:2011}, GMSD \cite{xue2014gradient}, SFF \cite{chang2013sparse}, SCQI \cite{bae2016novel}, ADD-GSIM \cite{gu2016analysis}, and SR-SIM \cite{zhang2012sr}.

The MTL framework has the following merits. Firstly, it is very convenient to generate distorted images given large-scale pristine reference images and to estimate their objective quality scores via FR-IQA methods, which motivated us to create the large distorted image dataset KADIS-700K. Therefore, training a deep CNN on large-scale images to address over-fitting becomes feasible. 
Secondly, although several FR-IQA methods have been proposed in the last decade, there is no single FR-IQA method that significantly outperforms the rest. One method may be superior for some distortion types, but inferior for others as different methods extract different features and use different pooling strategies. Therefore, learning to predict objective quality scores from multiple FR-IQA methods rather than from a single FR-IQA method by using a shared CNN body reduces the risk of over-fitting.  

\subsubsection{Loss function}
Mean Squared Error (MSE) loss is superior to Mean Absolute Error (MAE) loss for deep-learning-based IQA \cite{hosu2019koniq}.  
Therefore, we applied MSE as a loss function for each separate task. 
Let $Y=(y_1,\ldots,y_N)$ and $\hat{Y}=(\hat{y}_1,\ldots,\hat{y}_N)$ be ground truth scores and predicted scores, the MSE loss is defined as:
\begin{equation}\label{eq:mse}
    L_{\text{MSE}}(Y,\hat{Y}) = \frac{1}{N}\sum_{n=1}^N (y_n - \hat{y}_n)^2
\end{equation}

The loss function of the MTL framework is defined as:
\begin{equation}
    L_{\text{MTL}} = \sum_{k=1}^K \alpha_k L_{\text{MSE}}(Y_k,\hat{Y}_k), 
\end{equation}
where $Y_k$ and $\hat{Y}_k$ correspond to predicted objective scores and their corresponding ground truth objective scores of $k$-th FR-IQA metric. In our model, we assume each task has equal weight, namely $\alpha_k = 1/K$.

\subsubsection{Histogram equalization}

\begin{figure}[!htbp]
\centering
\begin{minipage}{.46\linewidth}
\centerline{\includegraphics[width=\textwidth]{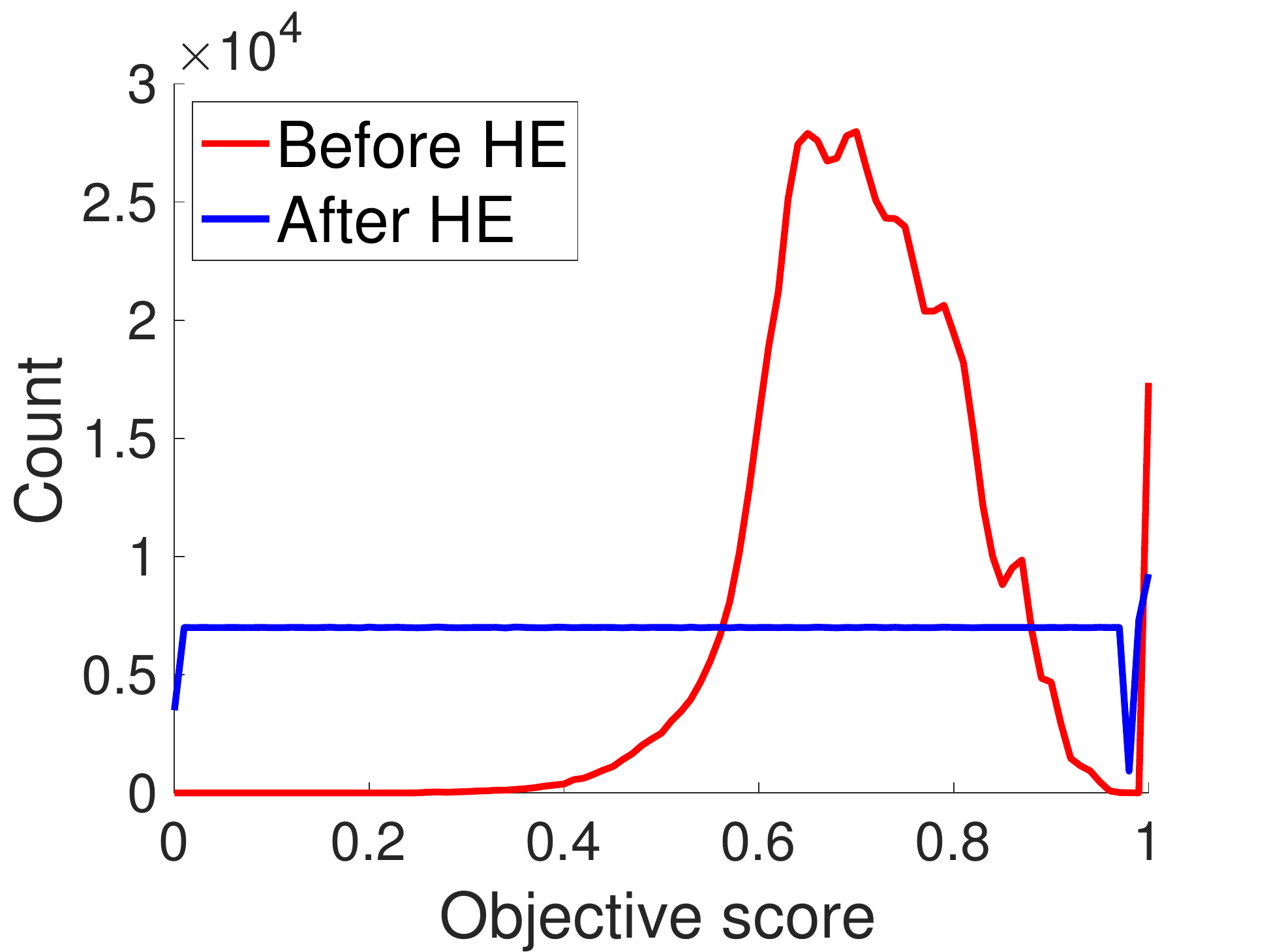}}
\centerline{(a) MDSI}
\end{minipage}%
\begin{minipage}{.46\linewidth}
\centerline{\includegraphics[width=\textwidth]{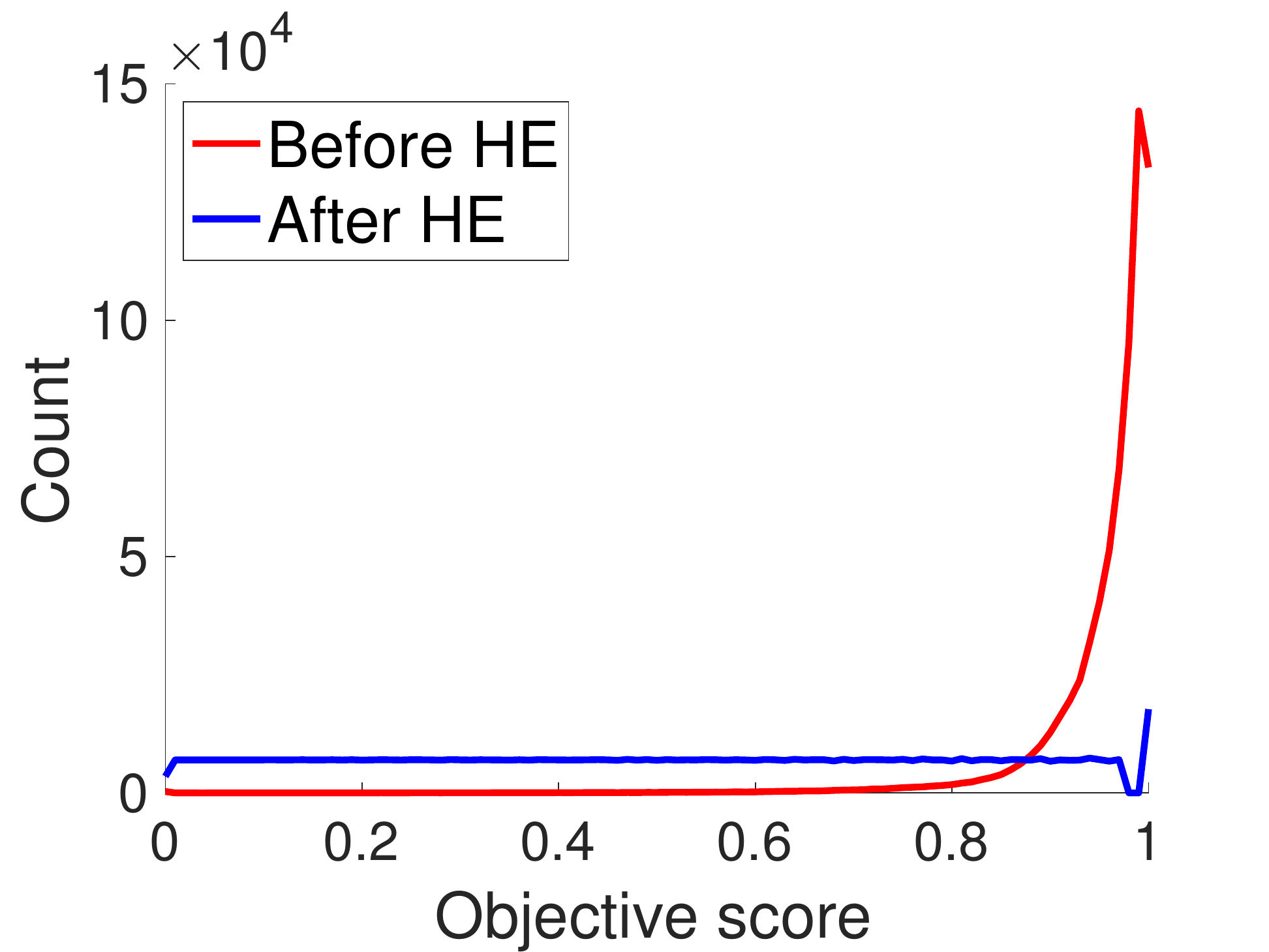}}
\centerline{(b) SFF}
\end{minipage}
\caption{An example of histogram equalization (HE) for FR-IQA scores. The original histograms of (a) MDSI and (b) SFF scores (red) in KADIS-700k are distributed in a narrow range, whereas both scores (blue) are uniformly distributed after HE.}
\label{fig:he_example}
\end{figure}

The performance of the MTL model is strongly dependent on the relative weighting between the loss of each task. To address this problem, one may tune the weights manually, which is a time-consuming and challenging process. A more principled way is to automatically assign the weights to each task according to some criterion, e.g., \cite{kendall2018multi,chen2018gradnorm,guo2018dynamic}. Different from the existing work, we applied histogram equalization \cite{Gonzalez:2006} to solve this problem.

In the MTL stage, the objective quality score distribution of FR-IQA methods are different from each other. 
Some FR-IQA methods produce values in a narrow range, resulting in a ``peaky'' distributions of scores over the images in our training set, e.g., Fig.~\ref{fig:he_example}(b). This results in different tasks having different learning speeds during training. After training a while, some tasks may already have achieved the optimal performance and start to over-fit, whereas the other tasks may still need more time to train. Consequently, the overall optimal performance across all tasks may never be reached. 
In order to better account for multiple peaks or other ab-normal score distribution shapes, we flattened the distributions by applying histogram equalization on the binned FR-IQA method scores. An example is illustrated in Fig.~\ref{fig:he_example}.



\subsection{Image representation by MLSP features}
\begin{figure}[!htbp]
\centering
\includegraphics[width=1\linewidth]{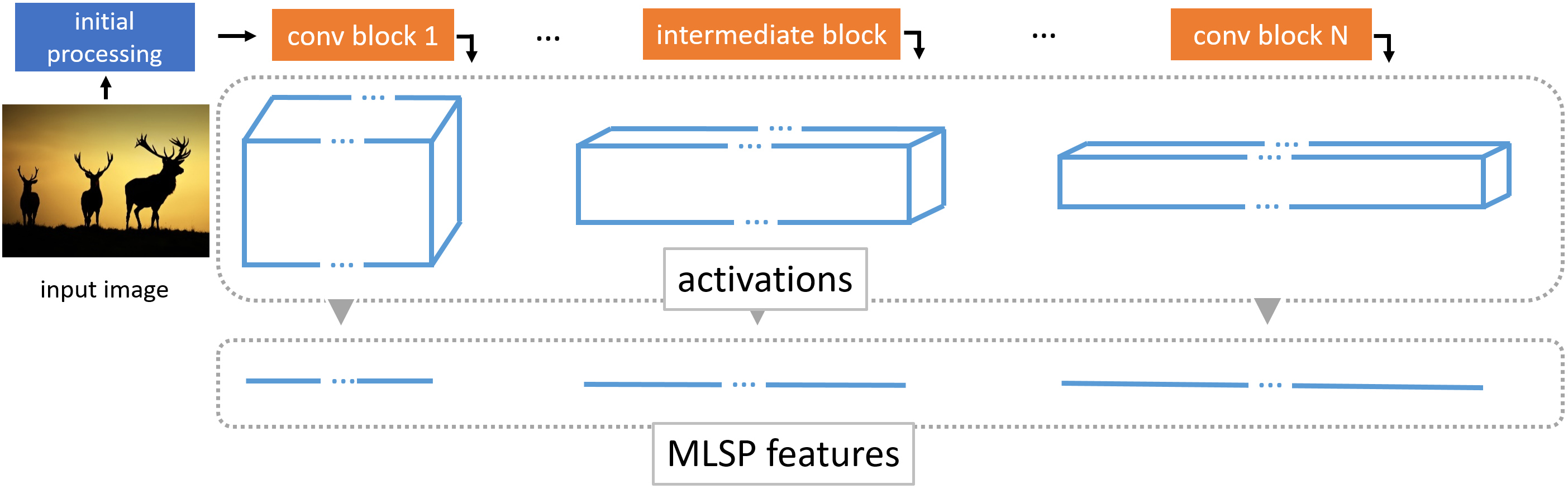}
\caption{Feature extraction from a fine-tuned CNN. Activation spatial sizes steadily decrease from the earlier convolution blocks to the later ones, while the channels dimension increases. In InceptionResNet-v2 there are 43 such blocks. The narrow 1$\times$1$\times$16928 MLSP features are extracted, by applying a Global Average Pooling operation on the activations of each block, and concatenating the results.}
\label{fig:MLSP_features}
\end{figure}

After training the MTL framework on a large dataset, one can fine-tune the pre-trained CNN on the benchmark IQA databases for predicting visual quality. This strategy is not recommended for best results, due to the following reasons:
\begin{itemize}
\item Many IQA databases are very small relative to the usual database training sizes used in state-of-the-art CNN models for object classification (1000 images per class), like the ones we use.
\item Retraining is expected to "forget" some of the useful information that was learnt in the first MTL stage of the InceptionResNetV2 network on KADIS-700k.
\item Images in IQA databases have various sizes and aspect ratios, which may be different from the ones used in the first fine-tuning stage; standardizing image size in all datasets means some information will be lost, as suggested by \cite{Hosu2019Effective}.
\end{itemize}
Training on Multi-Level Spatially Pooled (MLSP) features \cite{Hosu2019Effective} has been shown to work very well for Aesthetics Quality Assessment (AQA), and provided a solution to the problems above. As shown in Fig.~\ref{fig:MLSP_features}, we propose to use similar features for IQA:
\begin{itemize}
\item The features represent and ``lock-in'' the information embedded in the network resulting from the first fine-tuning stage.
\item Locked-in features do not allow to forget previously trained information and require much smaller prediction models, which may lead to better performance on existing benchmark datasets (small in size).
\item MLSP features have a fixed size, independent of input image size.
\end{itemize}

\subsection{Visual quality prediction}
\begin{table*}[!htbp]
\footnotesize
\caption{Benchmark IQA databases for performance evaluation.}
\label{tb:benchmarkdb}
\centering
\resizebox{1.0\textwidth}{!}{
\begin{tabular}{l r r c c r r c c} \hline
& &\multicolumn{1}{c}{No.\ of}&& No.\ of&\multicolumn{1}{c}{No.\ of}&\multicolumn{1}{c}{Ratings}&& \multicolumn{1}{c}{Subjective study} \\
Database &Content&distorted images &Distortion type&distortion types &rated images &per image &Score range&environment\\ \hline
LIVE \cite{sheikh:2006statistical}  &29&779& artificial & 5&779&23&[1,100]&lab \\
CSIQ \cite{larson:2010most} &30 &866&artificial& 6&866&5$\sim$7&[0,1]&lab \\
TID2013 \cite{ponomarenko:2015image}&25&3,000&artificial&24&3,000&9&[0,9]&lab \\
LIVE in the Wild \cite{ghadiyaram:2016massive}&1,169&1,169&authentic&N/A&1,169&175&[1,100]&crowdsourcing \\
KonIQ-10k \cite{hosu2019koniq}&10,073&10,073&authentic&N/A&10,073&120&[1,100]&crowdsourcing \\
KADID-10k &81&10,125&artificial&25&10,125&30&[1,5]&crowdsourcing \\
KADIS-700k &140,000&700,000&artificial&25&0&0&N/A&N/A \\
\hline
\end{tabular}
}
\vspace{-5pt}
\end{table*}


Once we have the MLSP feature representation (16928 dimensions) for each image, we could use a regression model such as Support Vector Regression (SVR) to non-linearly map these features to their corresponding subjective quality scores. However, training such classical regression models is time-consuming when the number of features is very high. Instead, we propose a shallow regression network for this task. The architecture is illustrated in Fig.~\ref{fig:vqp_arch}. 
Given the MLSP feature vector that represents a given image, four fully-connected (FC) layers connect to it: the first three layers have 2,048, 1,024, and 256 neurons. All the three FC layers use the rectified linear unit (ReLU) as the activation function and are followed by a dropout layer with rates of 0.25, 0.25, and 0.5, respectively, to reduce the risk of over-fitting. The output layer has one output neuron with a linear activation, which predicts visual quality scores. The same as the first stage, MSE loss (Eq.~(\ref{eq:mse})) was used for optimization. 

\section{Experiments}

\subsection{Datasets for evaluation}

In our experiments, we used KADIS-700k for feature learning via MTL and KADID-10k for visual quality prediction. Apart from evaluating our model on KADID-10k, we also evaluated it on five benchmark IQA databases, shown in Table~\ref{tb:benchmarkdb}. Overall, our model was evaluated on four artificially distorted IQA databases and two authentically distorted IQA databases.

Two widely used correlation metrics were applied to evaluate the performance in our experiments: the Pearson Linear Correlation Coefficient (PLCC) and the Spearman Rank Order Correlation Coefficient (SROCC). The range of both metrics is $[-1, 1]$, where a higher value indicates a better performance.
The 5-parameter logistic function from \cite{sheikh:2006statistical} was used to map between the objective and subjective scores so that PLCC was fairly measured.

\begin{figure}[t]
\centering
\includegraphics[width=0.47\textwidth]{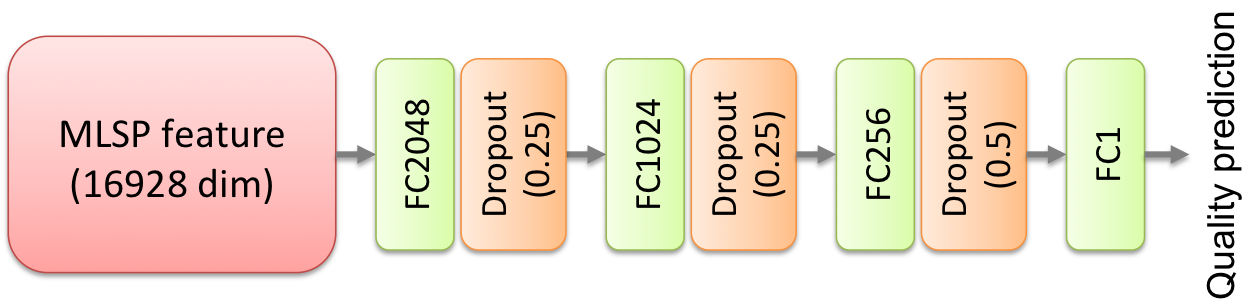}
\caption{The architecture for visual quality prediction in the second stage.}
\label{fig:vqp_arch}
\vspace{-5pt}
\end{figure}

\subsection{Training setup}
To make a fair and comprehensive comparison, our experiments were conducted as follows: In the MTL stage, KADIS-700k was divided into a training set of 600,000 images and a validation set of 100,000 images. There was no test set as it was unnecessary to evaluate the model at this stage. 
In the visual quality prediction stage, all databases were randomly split into a training set (60\%), a validation set (20\%), and a test set (20\%). All the divisions are on the basis of the content of reference images so there is no content overlapping between the three subsets. 

All the models and loss functions in our experiments were implemented using the Python Keras library with Tensorflow as a backend \cite{keras} and ran on two NVIDIA Titan Xp GPUs with a batch size of 64.
The InceptionResNetV2 CNN body was initialized with pre-trained weights from ImageNet \cite{deng2009imagenet}. The weights of FC layers were initialized with the method of He et al.\ \cite{henormal}. The ADAM \cite{kingma2014adam} optimizer was used for optimization, with a custom learning rate $\alpha$ and default parameters $\beta_1 =0.9$, $\beta_2 = 0.999$. 
In the MTL stage, we fixed $\alpha$ at $10^{-4}$; In the visual quality prediction stage, we used $\alpha = 10^{-1},10^{-2},\dots,10^{-5}$ and chose the value that gave the best validation result. 

Considering the large size of KADIS-700k, we only trained for 10 epochs in the MTL stage.
We trained for 30 epochs in the visual quality prediction stage. In the training process, we saved the best model that gave the minimum loss on the validation set and reported its performance on the test set. For visual quality prediction, we reported the \textit{median} values of the correlation metrics of 100 repetitions, the same criterion as other work, e.g., \cite{ye2012unsupervised}. 

\subsection{Evaluation on KADID-10k}


\subsubsection{Empirical analysis}

\begin{figure}[t]
\centering
\includegraphics[width=0.8\linewidth]{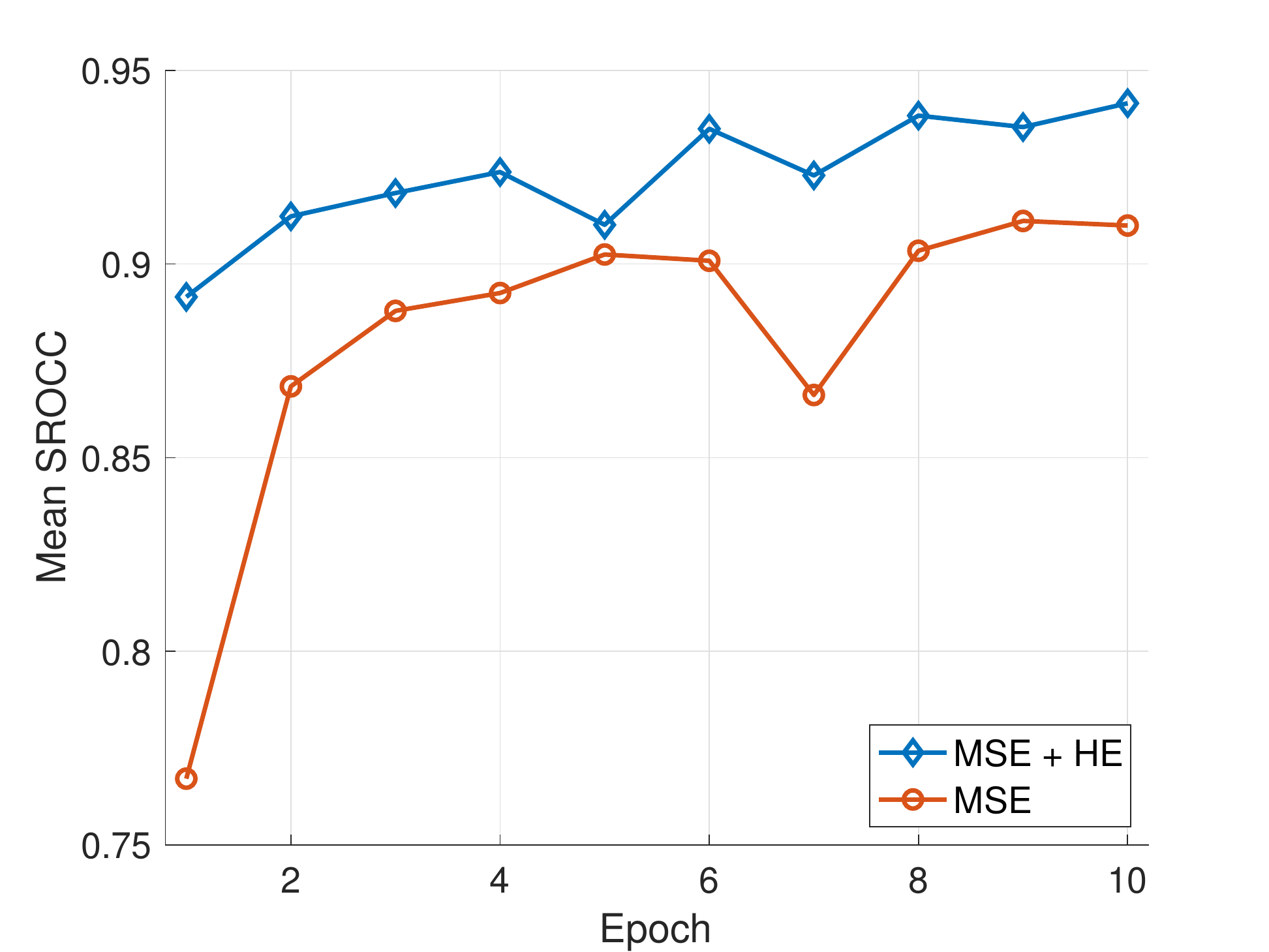}
\caption{Mean SROCC w.r.t.\ MSE loss with/without histogram equalization on the validation set of KADIS-700k.}
\label{fig:mtl_loss_he}
\end{figure}

We first investigated the effects of Histogram Equalization (HE) in the MTL stage.  We compared models trained with and without HE on objective score prediction. The mean SROCC of eleven quality prediction tasks on the validation set of KADIS-700k after every epoch is displayed in Fig.~\ref{fig:mtl_loss_he}. As can be seen, training with HE produces a more stable and better learning profile than without HE. 

\begin{figure}[t]
\centering
\includegraphics[width=0.9\linewidth]{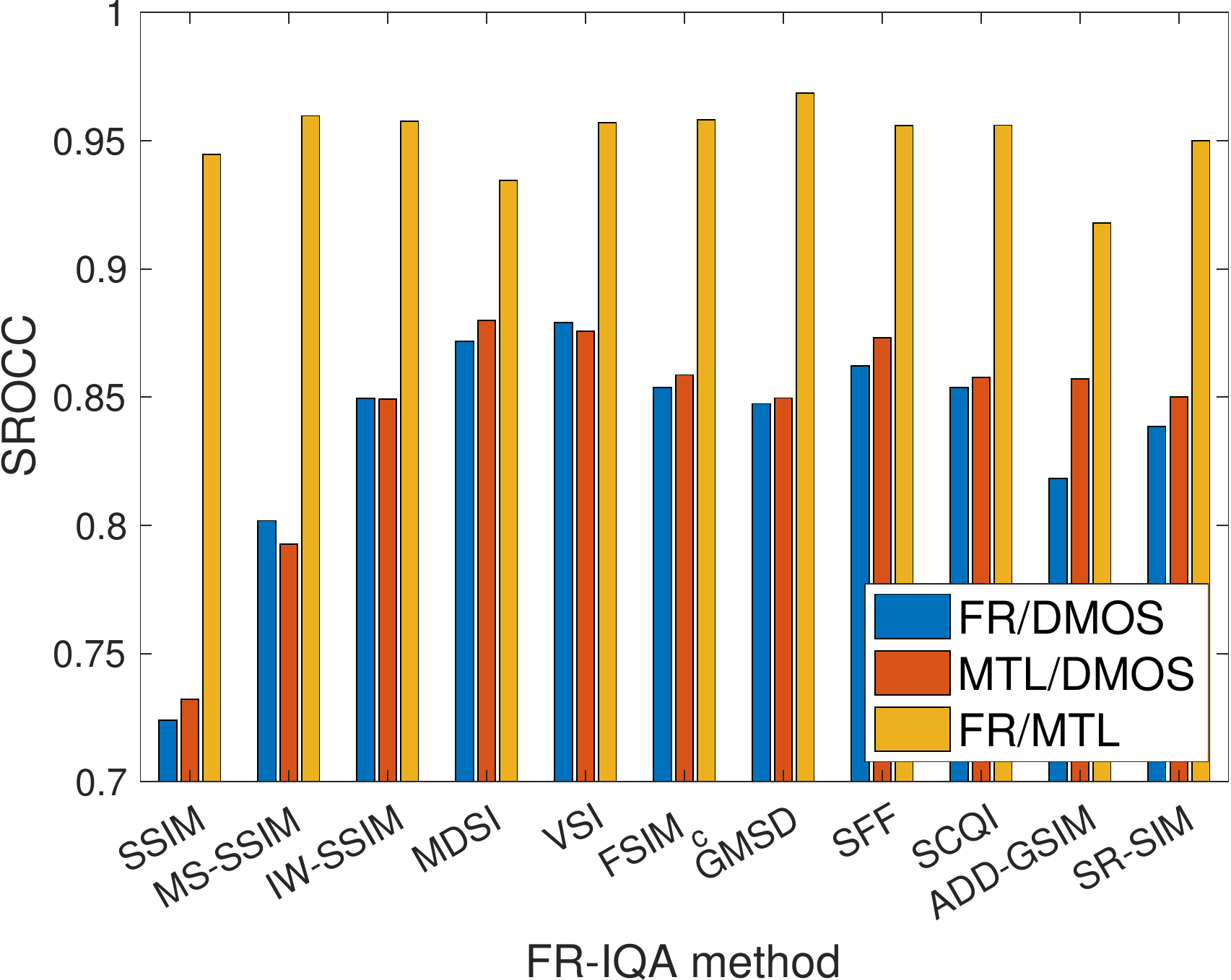}
\caption{Performance evaluation of the MTL model on KADID-10k. For each FR-IQA method, we compute three SROCCs: SROCC between objective scores of the FR-IQA method and DMOS (FR/DMOS), SROCC between objective scores predicted by MTL model and DMOS (MTL/DMOS), and SROCC between objective scores of the FR-IQA method and objective scores predicted by MTL model (FR/MTL).}
\label{fig:mtl_result}
\end{figure}

Given the MTL model that achieved the best mean SROCC on the validation test of KADIS-700k, we evaluated its performance on KADID-10k in the visual quality prediction stage, as shown in Fig.~\ref{fig:mtl_result}. Our MTL model has a very high correlation (over 0.92 SROCC)  w.r.t.\ objective scores estimated by each FR-IQA method.
Therefore, the quality predictions obtained by our MTL model, trained and tested on KADID-10k, are very similar to those obtained by the FR-IQA methods. This is the case even though our method works without the reference images.


Next, we compared the performance of MTL with a single-task learning strategy. We chose to predict objective scores of either the best FR-IQA metric, VSI, or the worst one, SSIM, according to its performance on KADID-10k, see Table \ref{tb:comp_kadid10k}. The training process for single-task learning was the same as that for MTL, except for replacing MTL modules by a single task module. The MLSP features extracted from the three learning strategies were used to predict visual quality. As a result, MTL obtained 0.936 SROCC on KADID-10k, outperformed the single learning strategy, which obtained 0.927 when predicting the VSI score and 0.923 when predicting the SSIM score.  

\begin{table}[t]
\centering
\begin{tabular}{l  c  c  c }
Scheme & PLCC   & SROCC \\ \hline
1. Fine-tune (ImageNet) &0.734 &0.731 \\
2. Fine-tune (KADIS-700k) &0.909 &0.908 \\
3. MLSP (ImageNet) & 0.712 & 0.706  \\
4. MLSP (KADIS-700k) &\bf 0.938 &\bf 0.936 \\ \hline
\end{tabular}
\caption{Performance comparison on KADID-10k with different learning schemes.}
\label{tb:perf_comp}
\end{table}

Finally, we compared the performance of the following four learning schemes on KADID-10k:

\begin{itemize}
\item[1] \textit{Fine-tune (ImageNet).}
In the first scheme, we replaced the output layer of InceptionResNetV2 CNN \cite{szegedy2017inception}, a softmax layer with 1,000 neurons, by a linear layer with one neuron. With the initialized weights that were pre-trained on ImageNet, the network was fined-tuned on KADID-10k for quality prediction.

\item[2] \textit{Fine-tune (KADIS-700k)}
The second scheme used the same architecture and the same initialized weights as the first scheme. However, it was first fine-tuned on KADIS-700k via MTL before it was fine-tuned on KADID-10k.

\item[3] \textit{MLSP (ImageNet)} In the third scheme, we trained a shallow regression network (Fig.~\ref{fig:vqp_arch}) for quality prediction based on MLSP features, which were extracted from a pre-trained network on ImageNet.

\item[4] \textit{MLSP (KADIS-700k)} The fourth scheme, which is used by our DeepFL-IQA model, trained the same regression network as the third scheme. However, its MLSP features were extracted from fine-tuned weights on KADIS-700k instead of ImageNet.
\end{itemize}

\begin{figure*}[t]
\centering
\begin{minipage}{0.225\linewidth}
\centerline{\includegraphics[width=\textwidth]{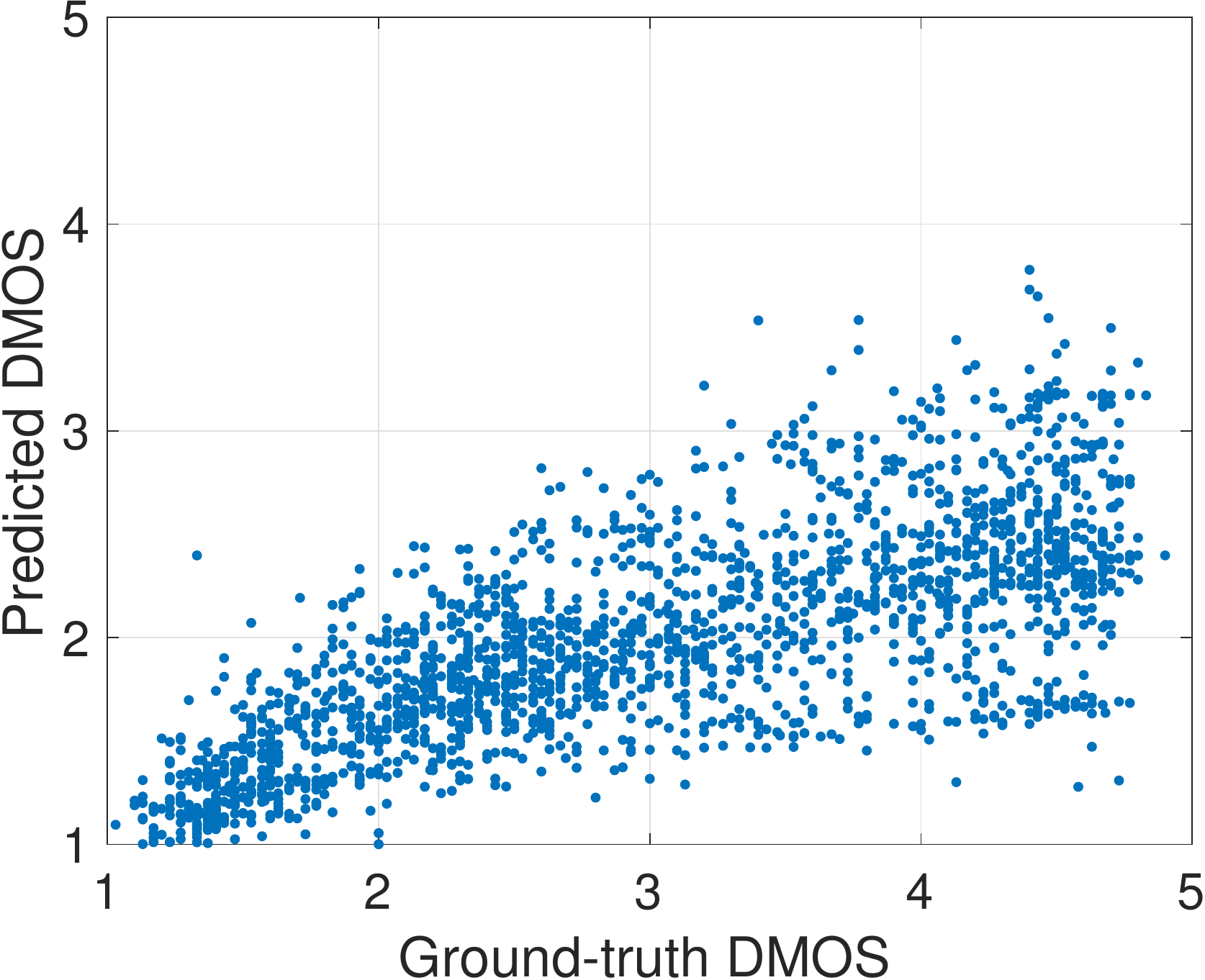}}
\centerline{(a) Fine-tune (ImageNet)}
\end{minipage}
\begin{minipage}{0.225\linewidth}
\centerline{\includegraphics[width=\textwidth]{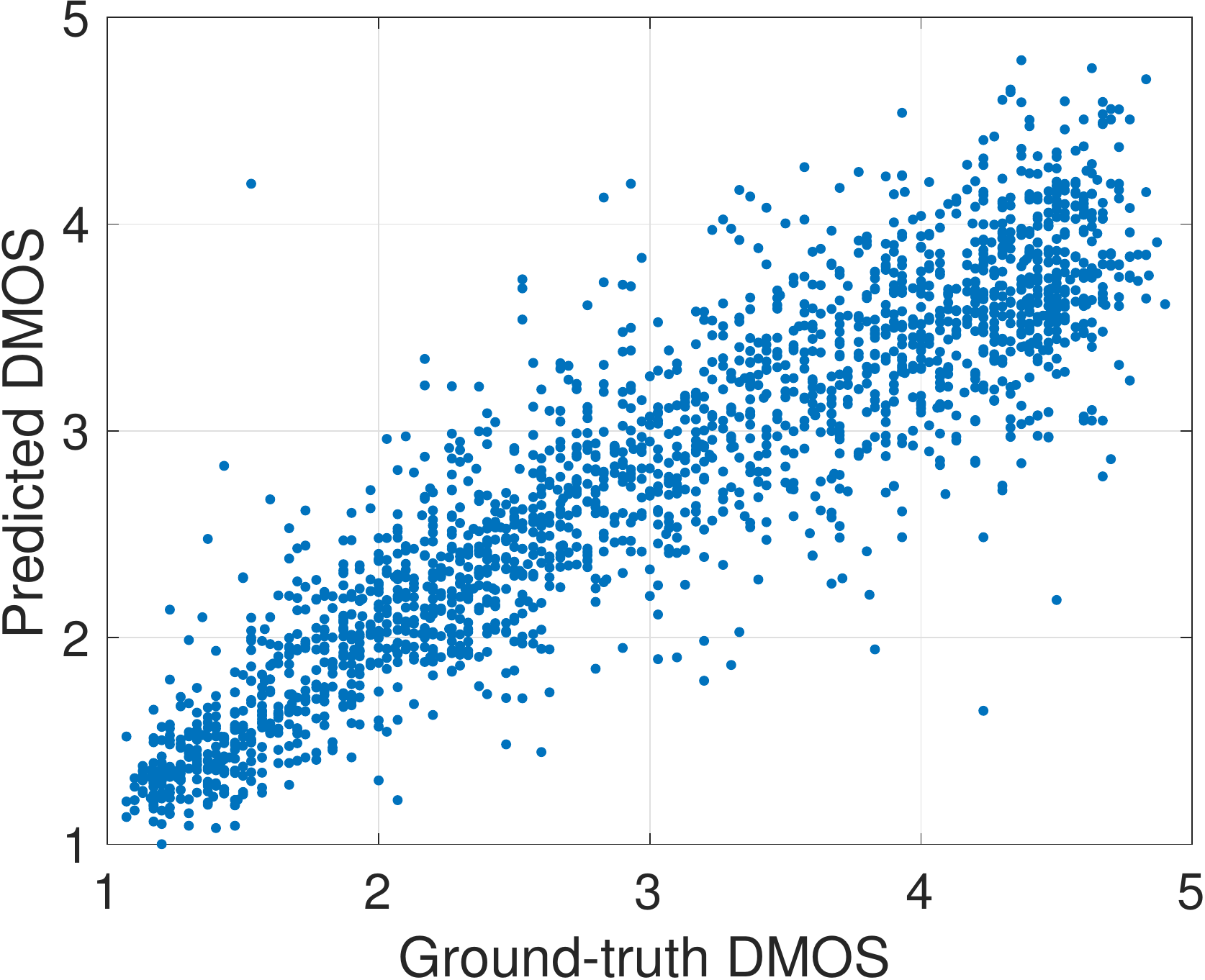}}
\centerline{(b) Fine-tune (KADIS-700k)}
\end{minipage}
\begin{minipage}{0.225\linewidth}
\centerline{\includegraphics[width=\textwidth]{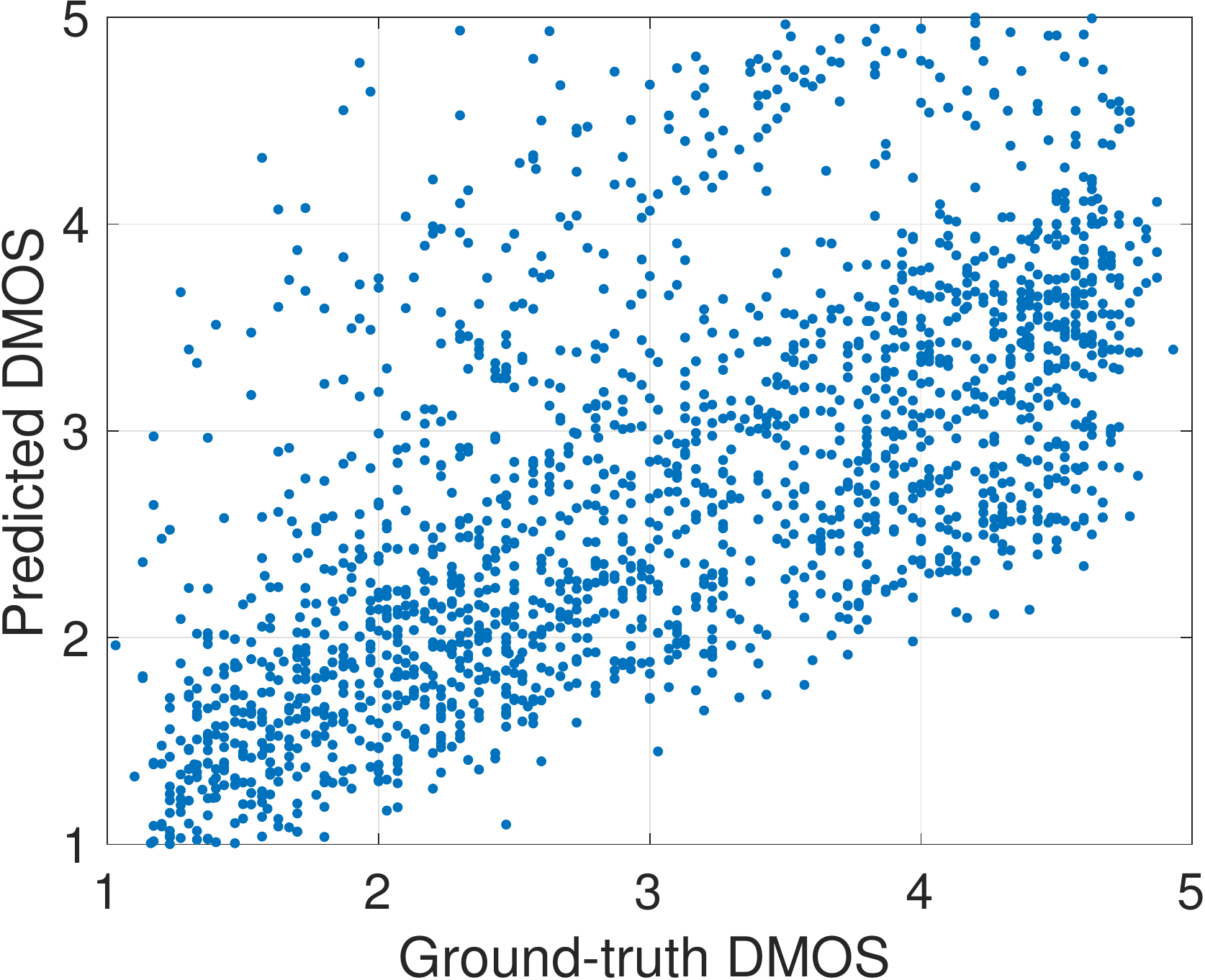}}
\centerline{(c) MLSP (ImageNet)}
\end{minipage}
\begin{minipage}{0.225\linewidth}
\centerline{\includegraphics[width=\textwidth]{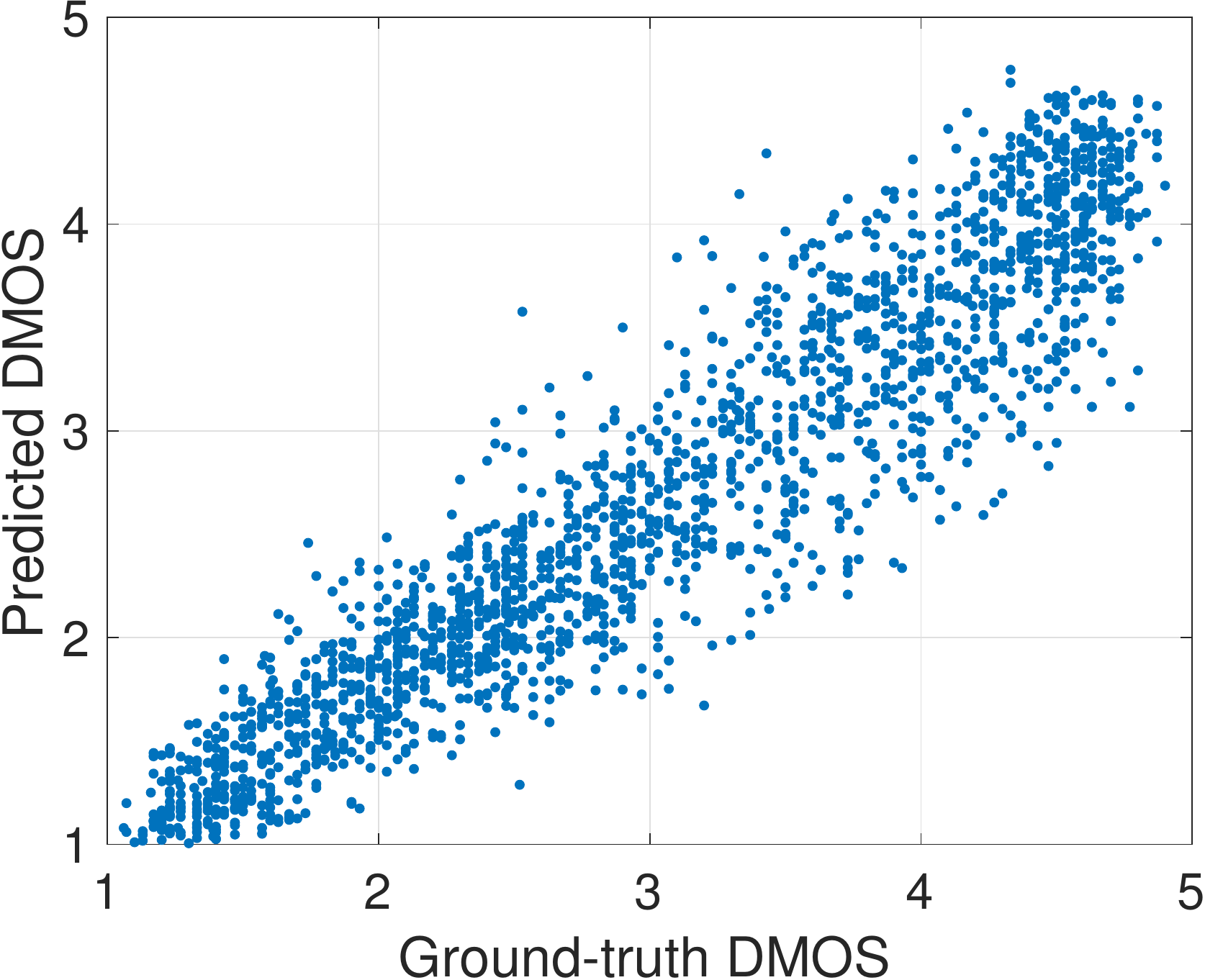}}
\centerline{(d) DeepFL-IQA}
\end{minipage}
\caption{Scatter plots of predicted DMOS versus ground truth DMOS on the KADID-10k database. 
}
\label{fig:kadid10k_scatter_plot}
\vspace{-5pt}
\end{figure*}

Table~\ref{tb:perf_comp} shows the performance of the four schemes on KADID-10k, and their scatter plots are presented in Fig.~\ref{fig:kadid10k_scatter_plot}. The learning scheme we used, i.e., transfer learning from the image classification domain (ImageNet) to the IQA domain (KADIS-700k), together with MLSP feature learning, outperforms the remaining schemes. 

\subsubsection{Comparison with previous work}

We compared our DeepFL-IQA with some FR-IQA and NR-IQA methods. 
Since the results of deep-learning-based approaches are difficult to reproduce without available source code, here we only compared conventional NR-IQA methods with available source code.
Seven conventional NR-IQA methods were collected, including BIQI \cite{Moorthy:2010}, BLIINDS-II \cite{blind2}, BRISQUE \cite{bris}, CORNIA \cite{ye2012unsupervised}, DIIVINE \cite{Moorthy:2011}, HOSA \cite{hosa}, and SSEQ \cite{Liu:2014b}. We used an SVR (RBF kernel) to train and predict quality from the  handcrafted features extracted by the NR-IQA methods. 


Table~\ref{tb:comp_kadid10k} shows the comparison results on the KADID-10k database. 
Although the best FR-IQA method, VSI, achieves 0.879 SROCC, which is much better than conventional NR-IQA methods, there is still room for improvement. 
DeepFL-IQA improves the SROCC by around 0.06 compared to VSI and around 0.3 compared to conventional feature-learning-based methods, e.g., CORNIA and HOSA. 

Table~1 (Supplementary Material) compares the SROCC in each individual distortion. As can be seen, FR-IQA methods still outperform DeepFL-IQA in most distortions. This is an expected result, considering that our MLSP features are learned from objective scores of FR-IQA methods, and the method is NR-IQA only. However, MTL makes MLSP features generalize better than single-task learning strategy, thus achieves good results across all distortions. As a result, the overall performance is better than that of any single FR-IQA method.

\begin{table}[t]
\centering
\begin{tabular}{c|l| c c c c c c c c } \midrule[1.0pt]
&Method & PLCC & SROCC  \\ \midrule[1.0pt] 
\multirow{11}{*}{\rotatebox{90}{FR-IQA}}&SSIM & 0.723&0.724 \\
& MS-SSIM &0.801& 0.802  \\
& IW-SSIM &0.846&0.850  \\ 
&MDSI & 0.873 &0.872  \\ 
&VSI &\bf 0.878 &\bf 0.879  \\
&FSIM$_{\text{c}}$ & 0.851 & 0.854 \\
&GMSD & 0.847 & 0.847  \\
&SFF & 0.862 & 0.862\\
&SCQI & 0.853 &0.854  \\
&ADD-GSIM &0.817 &0.818  \\
&SR-SIM & 0.834 & 0.839 \\\hline 
\multirow{10}{*}{\rotatebox{90}{NR-IQA}}& BIQI & 0.445 &0.425  \\ 
&BLIINDS-II &0.557 &0.537 \\
&BRISQUE &0.567 &0.528 \\
&CORNIA &0.558 &0.516 \\
&DIIVINE &0.521 &0.479\\
&HOSA &0.653 &0.618  \\ 
&SSEQ &0.532 &0.487  \\
& DeepFL-IQA &\bf 0.938 &\bf 0.936 \\ 
\midrule[1.0pt]
\end{tabular}
\caption{Performance comparison on KADID-10k database. Top FR-IQA and NR-IQA models are highlighted.}
\label{tb:comp_kadid10k}
\end{table}


\subsection{Generalization of deep features}
To investigate the generalization of deep-features learned in the MTL stage, we carried out the following experiments. Firstly, in the second stage, we trained an SVR instead of a shallow regression neural network. Secondly, we evaluated DeepFL-IQA on other benchmark databases that have several distortion types not available in KADIS-700k. Last but not least, we carried out a cross-database evaluation on all the databases.

\subsubsection{The power of deep-features}
By training an SVR in the visual quality prediction stage, we obtained an SROCC of 0.930 and PLCC of 0.932, slightly worse than using the shallow regression network (0.936 SROCC and 0.938 PLCC). This proves that the deep-features learned in the MTL stage work well for IQA independently of the regression model. 


\begin{table}[t]
\centering
\begin{tabular}{l | c c c c  } \midrule[1.0pt]
 & \multicolumn{2}{c}{KonIQ-10k}& \multicolumn{2}{c}{LiveitW} \\
Method & PLCC & SROCC & PLCC & SROCC \\ \midrule[1.0pt] 
BIQI&0.707 &0.680 &0.546 &0.521  \\ 
BLIINDS-II &0.613 &0.603 &0.435 &0.394\\
BRISQUE &0.716 &0.705 &0.639 &0.603  \\
CORNIA &0.774 &0.755 &0.691 &0.647\\
DIIVINE &0.714 &0.701 &0.602 &0.581\\
HOSA &0.806 &0.792 &0.673 &0.650  \\ 
SSEQ &0.590 &0.575 &0.439 &0.406\\
MLSP (ImageNet)&\bf 0.924 & \bf 0.913 &\bf 0.842 &\bf 0.814\\ 
DeepFL-IQA &0.887 &0.877 & 0.769 & 0.734\\ 
\midrule[1.0pt]
\end{tabular}
\caption{Performance comparison on the two authentically distorted benchmark databases. Top IQA model is highlighted.}
\label{tb:mlsp_comp_wild_db}
\vspace{-5pt}
\end{table}

\subsubsection{Evaluation on other benchmark databases}

We compared DeepFL-IQA with the state-of-the-art IQA methods on five other benchmark databases, using the MTL model, pre-trained on KADIS-700k.
Specifically, LIVE, CSIQ, and TID2013 are artificially distorted databases we used; Live in the Wild (LiveitW), and KonIQ-10k are authentically distorted databases we used.  
For artificially distorted databases, both FR-IQA and NR-IQA methods were evaluated, whereas only NR-IQA methods were evaluated in authentically distorted databases as reference images do not exist. The NR-IQA methods are grouped into conventional and deep-learning-based. The results for the deep-learning-based approaches are difficult to reproduce. Therefore, we can only report the performance numbers as reported in their respective papers. 

\begin{table*}[t]
\centering
\begin{tabular}{c|l| c c c c c c | c c} \midrule[1.0pt]
& & \multicolumn{2}{c}{LIVE} & \multicolumn{2}{c}{CSIQ}& \multicolumn{2}{c|}{TID2013}& \multicolumn{2}{c}{Weighted Avg.}  \\
&Method  & PLCC & SROCC & PLCC & SROCC & PLCC & SROCC & PLCC & SROCC \\ \midrule[1.0pt] 
\multirow{11}{*}{\rotatebox{90}{FR-IQA}}&SSIM &0.945 &0.948&0.861 &0.876  & 0.691 & 0.637 & 0.766 & 0.735 \\
& MS-SSIM  & 0.949 & 0.951&0.899 &0.913 &0.833 &0.786 &0.865 &0.838 \\
& IW-SSIM&0.952&0.957& 0.914 &0.921 &0.832 &0.778 &0.868 &0.835 \\ 
&MDSI  &  0.966 & 0.967 &0.953 &0.957&\bf 0.909 &0.890 &\bf 0.927 &0.916 \\ 
&VSI  &0.948&0.952&0.928 &0.942 &0.900&0.897 &0.913 &0.915 \\
&FSIM$_{\text{c}}$  & 0.961 &0.965 & 0.919 &0.931 &0.877 &0.851 &0.899 &0.885 \\
&GMSD  & 0.960 &0.960 &0.954 &0.957 & 0.859 &0.804 &0.894 &0.859 \\
&SFF  &0.963 &0.965 &\bf 0.964 & \bf 0.963&0.870&0.851 &0.903 &0.891\\
&SCQI  &0.934 &0.941 & 0.927 &0.943 &0.907 &\bf 0.905 &0.915 &\bf 0.918  \\
&ADD-GSIM  & \bf 0.967 & \bf 0.968 &0.934 &0.942 &0.880 &0.829 &0.905 &0.874  \\
&SR-SIM  & 0.955 &0.961 &0.925 &0.932 &0.866 &0.807 &0.892 &0.857 \\
\hline 
\multirow{15}{*}{\rotatebox{90}{NR-IQA}}& BIQI  &0.764 &0.748 &0.553 &0.523 &0.519 &0.399 &0.567 &0.482 \\ 
&BLIINDS-II  &0.949 &0.946 &0.803 &0.760 &0.563 &0.557 &0.672 &0.661\\
&BRISQUE &0.951 &0.937 &0.775 &0.691 &0.480 &0.415 &0.615 &0.556\\
&CORNIA  &0.961 &0.951 &0.796 &0.721 &0.705 &0.612 &0.766 &0.690\\
&DIIVINE  &0.909 &0.900 &0.789 &0.754 &0.687 &0.598 &0.744 &0.769 \\
&HOSA  &0.961 &0.948 &0.834 &0.783 &0.762 &0.678 &0.809 &0.744 \\ 
&SSEQ  &0.905 &0.888 &0.790 &0.707 &0.660 &0.579 &0.726 &0.656 \\
\cline{2-10}
&DIQaM-NR \cite{bosse2018deep}&0.972&0.960 &-- &-- &0.855&0.835 & -- &-- \\
&RankIQA \cite{Liu_2017_ICCV} &0.982&0.981&0.960&0.947&0.799 &0.780 & 0.860 &0.845 \\
&MEON \cite{ma2018end} & -- &-- &0.944 &0.932 & -- &0.808 &-- &--\\
&BPSOM-MD \cite{pan2018blind} &0.955 &0.967 &0.860 &0.904 &0.879 &0.863 &0.889 &0.888\\ 
&BLINDER \cite{gao2018blind}  & 0.959 &0.966 &\bf 0.968 &\bf 0.961 &0.819 &0.838 &0.871 &0.883  \\
&NSSADNN \cite{yan2019naturalness}&\bf0.984&\bf 0.986&0.927&0.893 &\bf 0.910&0.844 &\bf 0.926 &0.877 \\
&Hallucinated-IQA \cite{lin2018hallucinated} & 0.982 & 0.982 &0.910 &0.885 & 0.880 &\bf 0.879 & 0.903 &\bf 0.898  \\
&DeepFL-IQA  &0.978 &0.972 &0.946 &0.930 &0.876 &0.858 & 0.907 &0.891\\ 
\midrule[1.0pt]
\end{tabular}
\caption{Performance comparison on the three artificially distorted benchmark databases. All the FR-IQA and conventional NR-IQA methods are re-implemented. For deep-learning-based IQA methods, we list the performance numbers as reported in their respective papers due to the difficulty for re-implementation. Top FR-IQA and NR-IQA models are highlighted.}
\label{tb:mlsp_comp_ad_db}
\end{table*}

\begin{table*}[t]
\small
\centering
\begin{tabular}{r l |c c c c c c}
\midrule[1.0pt]
& & \multicolumn{6}{c}{Test on} \\
& & KADID-10k& LIVE & CSIQ & \multicolumn{1}{c|}{TID2013} & KonIQ-10k &LiveitW \\ \midrule[1.0pt]
\multirow{6}{*}{\rotatebox{90}{Train on}} & KADID-10k & \bf 0.936 &0.894  &0.849  &\multicolumn{1}{c|}{0.710}  &0.474  &0.414  \\
 & LIVE &0.689 & \bf 0.972 &0.831 &\multicolumn{1}{c|}{0.577} &0.563  &0.454  \\
 & CSIQ &0.692  &0.942  & \bf 0.930 &\multicolumn{1}{c|}{0.636}  &0.282  &0.312  \\
 & TID2013 &0.698  &0.780  &0.816  & \multicolumn{1}{c|}{\bf 0.858} &0.386  &0.285  \\ \cline{2-8}
 & KonIQ-10k &0.737  &0.920  &0.746 &\multicolumn{1}{c|}{0.599}  & \bf 0.877 &0.704  \\
 & LiveitW &0.440  &0.651  &0.577 &\multicolumn{1}{c|}{0.384}  &0.711  &\bf 0.734 \\
 \midrule[1.0pt]
\end{tabular}
\caption{Cross database evaluation of DeepFR-IQA (SROCC) on six benchmark databases. The models are trained on one database and tested on the rest of entire databases. For comparison, the results on test set from the same database are also listed in bold.}
\label{tb:cross-test}
\vspace{-10pt}
\end{table*}

Table~\ref{tb:mlsp_comp_ad_db} shows the performance comparison on three widely used artificially distorted benchmark databases, i.e., LIVE, CSIQ, and TID2013. 
Although there are distortions in the three databases that were not learned in the MTL stage, e.g., fast fading in LIVE and global contrast decrements in CSIQ, DeepFL-IQA significantly outperformed conventional NR-IQA methods and obtained results comparable to other deep-learning-based methods, even though most of them were trained end-to-end given the benchmark databases. Overall, DeepFL-IQA achieves the second-best weighted average PLCC and SROCC among NR-IQA methods.
This further supports that our pre-trained MTL model on KADIS-700k generalizes well. 
We compared the SROCC on 24 distortions of TID2013 in Table~2 (Supplementary Material). 
As only 13 out of 24 distortions were included in KADIS-700k, the performance could be further improved when more distortions are trained in the MTL stage.


Table~\ref{tb:mlsp_comp_wild_db} presents the results on two authentically distorted databases, namely KonIQ-10k and LiveitW. 
Although DeepFL-IQA still outperformed conventional NR-IQA methods on both databases, it performed worse than the third learning scheme, namely, training MLSP features that were extracted from a pre-trained network on ImageNet rather than KADIS-700k. This demonstrates that the distortions in artificially distorted databases are very different from the distortions in the wild, where the image quality is more content dependent, and the predominant distortions are a combination of multiple distortions.

\subsubsection{Cross database test}
To further evaluate the generalization ability of our DeepFL-IQA model, we carried out a cross-database evaluation on six benchmark databases. Table~\ref{tb:cross-test} presents the results, where the models were trained on one database (60\% training, 20\% validation, and 20\% testing) and tested on the remaining databases. Obviously, due to the reasons such as noise in subjective scores, types of distortions, and image content, the performance decreased when conducting a cross-test evaluation. In particular, the performance decreased sharply when training on artificially distorted databases and testing on authentically distorted databases, or vice versa. This confirms that the distortions in these two broader types of databases are very different. Nonetheless, increasing the size of the dataset and number of types of distortions is prone to boost performance, e.g., by using KADID-10k and KonIQ-10k.

\section{Conclusion}

IQA is missing truly massive datasets for deep learning. We introduced such a dataset for weakly supervised learning, consisting of 700,000 (KADIS-700k) artificially degraded images without subjective scores, and 10,125 images of a similar kind that have been subjectively scored (KADID-10k).

Our collection of pristine images (KADIS-140k), is 30 times as large as the largest existing, the Waterloo dataset pristine images. We distorted images both by reproducing distortions from the literature (TID2013) and introducing new ones that relate to predominant defects in the wild. Our artificially distorted dataset (KADIS-700k), is 7 times as large as the entire Waterloo dataset. It can be further extended with the source code we released. 
Moreover, our subjectively scored dataset (KADID-10k) is reliably annotated, and 3 times larger than the best existing, TID2013.
The large datasets facilitate to develop superior deep learning models.

Based on our datasets KADIS-700k and KADID-10k, we proposed a deep model DeepFL-IQA. We applied an MTL framework to predict objective scores of eleven FR-IQA metrics, given only the distorted images in KADIS-700k. As the model is trained on a large dataset with content and distortion diversity, it has an excellent generalization. Rather than fine-tuning on small IQA databases, we extracted MLSP features for visual quality prediction, which allows input images with arbitrary aspect ratios and resolution. Evaluated on KADID-10k, our model improved around 0.2 SROCC compared to fine-tuned InceptionResNetV2 model on weights pre-trained from ImageNet and improved around 0.06 SROCC compared to best FR-IQA metric. Compared to other deep IQA models, DeepFL-IQA has several merits, such as (1) ease of implementation, (2) fast training and testing on other databases after MTL training, and (3) good generalization to unobserved distortions and other databases. 
Last but not least, we found the features learned from artificially distorted images are not suitable for IQA on authentically distorted images.

\section*{Acknowledgment}
Funded by the Deutsche Forschungsgemeinschaft (DFG, German Research Foundation) -- Project-ID 251654672 -- TRR 161 (Project A05).

\bibliographystyle{IEEEtran}
\bibliography{refs}

\end{document}


\maketitle

\section{Additional tables}

\begin{table*}[!htbp]
\centering
\resizebox{0.95\textwidth}{!}{
\begin{tabular}{c|l| c c c c c c c c c c c c c} \midrule[1.0pt]
&Method & \#01 & \#02 & \#03 & \#04 & \#05 & \#06 & \#07 & \#08 & \#09 & \#10 & \#11 & \#12& \#13 \\ \midrule[1.0pt] 
\multirow{11}{*}{\rotatebox{90}{FR-IQA}}&SSIM &0.949 &0.884 &0.940 &0.790 &0.696 &0.830 &0.522 &0.854 &0.928 &0.834 &0.839 &0.899 &0.878 \\
& MS-SSIM &0.950 &0.890 &0.940 &0.855 &0.746 &0.850 &0.595 &0.904 &0.934 &0.858 &0.852 &0.907 &0.879 \\
& IW-SSIM &\bf 0.960 &\bf 0.935 &\bf 0.961 &0.877 &\bf 0.783 &0.861 &0.495 &0.873 &0.948 &0.886 &0.854 &0.908 &0.861 \\ 
&MDSI &0.949 &0.901 &0.944 &0.860 &0.685 &0.876 &\bf 0.623 &0.906 &0.941 &0.899 &0.902 &0.935 &0.887 \\ 
&VSI &0.954 &0.911 &0.951 &\bf 0.917 &\bf 0.783 &0.870 &0.597 &0.934 &0.940 &0.886 &\bf 0.912 &\bf 0.944 &\bf 0.908 \\
&FSIM &0.954 &0.906 &0.950 &0.911 &0.697 &0.884 &0.609 &0.905 &\bf 0.951 &\bf 0.912 &0.841 &0.899 &0.870 \\
&GMSD &0.955 &0.909 &0.949 &0.863 &0.687 &0.868 &0.511 &0.882 &0.942 &0.890 &0.906 &0.938 &0.858 \\
&SFF  &0.952 &0.900 &0.945 &0.820 &0.657 &\bf 0.887 &0.550 &0.867 &0.947 &0.886 &0.899 &0.929 &0.879 \\
&SCQI &0.936 &0.864 &0.922 &0.898 &0.689 &0.839 &\bf 0.623 &\bf 0.935 &0.917 &0.842 &0.890 &0.916 &0.887 \\
&ADD-GSIM &0.956 &0.913 &0.947 &0.877 &0.729 &0.873 &0.470 &0.877 &0.943 &0.885 &0.882 &0.926 &0.864 \\
&SR-SIM &0.956 &0.928 &0.952 &0.884 &0.760 &0.878 &0.489 &0.879 &\bf 0.951 &0.899 &0.897 &0.931 &0.897 \\ \hline 
\multirow{9}{*}{\rotatebox{90}{NR-IQA}}& BIQI &0.856 &0.740 &0.558 &0.260 &0.137 &0.559 &-0.083 &0.154 &0.531 &0.284 &0.631 &0.625 &0.629 \\
&BLIINDS-II &0.863 &0.777 &0.492 &0.617 &0.123 &0.505 &0.204 &0.623 &0.651 &0.792 &0.658 &0.782 &0.644 \\
&BRISQUE &0.835 &0.819 &0.687 &0.512 &0.095 &0.709 &0.214 &0.396 &0.571 &0.756 &0.760 &0.779 &0.550 \\
&CORNIA &0.883 &0.883 &0.802 &0.277 &0.047 &0.648 &0.205 &0.289 &0.621 &0.757 &0.609 &0.720 &0.745 \\
&DIIVINE &0.779 &0.796 &0.625 &0.443 &0.235 &0.448 &0.007
&0.262 &0.565 &0.600 &0.613 &0.699 &0.591 \\
&HOSA &0.780 &0.829 &0.671 &0.453 &0.202 &0.639 &0.116 &0.409 &0.559 &0.746 &0.594 &0.631 &0.781 \\ 
&SSEQ &0.779 &0.717 &0.508 &0.468 &-0.029 &0.543
&0.055 &0.270 &0.498 &0.733 &0.790 &0.802 &0.770 \\ 
&DeepFL-IQA&\bf 0.934 &\bf 0.907 &\bf 0.927 &\bf 0.912 &\bf 0.836 &\bf 0.883 & \bf 0.603 &\bf 0.920 &\bf 0.932 &\bf 0.897 &\bf 0.920 &\bf 0.948 &\bf 0.892 \\ \midrule[1.0pt]
&Method & \#14 & \#15 & \#16 & \#17 & \#18 & \#19 & \#20 & \#21 & \#22 & \#23 & \#24 & \#25 &\multicolumn{1}{|c}{Overall} \\ \midrule[1.0pt]
\multirow{11}{*}{\rotatebox{90}{FR-IQA}}&SSIM  &0.893 &0.880 &0.864 &0.895 &0.662 &0.797 &0.426 &0.479 &0.727 &\bf 0.588 &0.879 &0.556  & \multicolumn{1}{|c}{0.724}       \\ & MS-SSIM &0.893 &0.917 &0.908 &0.894 &0.697 &0.922 &0.435 &0.477 &0.850 &0.578 &0.877 &0.766 &\multicolumn{1}{|c}{0.802} \\
& IW-SSIM &0.914 &\bf 0.918 &0.916 &\bf 0.897 &\bf 0.735 &\bf 0.960 &0.564 &0.564 &\bf 0.878 &0.541 &0.887 &\bf 0.814 & \multicolumn{1}{|c}{0.850}\\
&MDSI & 0.936 &0.910 &0.902 &0.888 &0.687 &0.951 &0.449 &0.514 &0.784 &0.546 &0.890 &0.758 &  \multicolumn{1}{|c}{0.872}     \\ 
&VSI  &\bf 0.946 &0.871 &0.909 &\bf 0.897 &0.694 &0.950 &0.472 &0.563 &0.692 &0.462 &\bf 0.916 &0.766  & \multicolumn{1}{|c}{\bf 0.879}     \\
&FSIM &0.917 &0.904 &0.917 &0.895 &0.689 &0.944 &0.466 &0.497 &0.812 &0.581 &0.889 &0.779 & \multicolumn{1}{|c}{0.854}     \\
&GMSD &0.935 &0.867 &\bf 0.918 &0.893 &0.642 &0.947 &0.409 &0.518 &0.785 &0.568 &0.874 &0.729 & \multicolumn{1}{|c}{0.847}     \\
&SFF  &0.931 &0.917 &0.911 &0.877 &0.692 &0.951 &0.452 &0.558 &0.716 &0.438 &0.908 &0.742 & \multicolumn{1}{|c}{0.862}      \\
&SCQI &0.943 &0.810 &0.904 &0.884 &0.668 &0.914 &0.368 &0.465 &0.741 &0.510 &0.840 &0.684 & \multicolumn{1}{|c}{0.854}      \\
&ADD-GSIM &0.920 &0.895 &0.876 &0.878 &0.675 &0.905 &0.462 &0.539 &0.759 &0.560 &0.894 &0.605 & \multicolumn{1}{|c}{0.818}  \\
&SR-SIM &0.935 &0.897 &0.916 &0.885 &0.682 &0.959 &\bf 0.571 &\bf 0.567 &0.684 &0.480 &0.905 &0.783 & \multicolumn{1}{|c}{0.839}    \\ \hline
\multirow{9}{*}{\rotatebox{90}{NR-IQA}}& BIQI & 0.706 &0.713 &0.228 &0.175 &0.063 &0.742 &0.227 &0.330 &0.369 &0.157 &0.614 &0.202 & \multicolumn{1}{|c}{0.425}  \\
&BLIINDS-II &0.672 &0.788 &0.450 &0.427 &0.212 &0.857 &0.076 &0.481 &0.560 &0.135 &0.653 &0.140 &\multicolumn{1}{|c}{0.537} \\
&BRISQUE &0.782 &0.830 &0.504 &0.439 &0.311 &0.737 &0.260 &0.623 &0.312 &0.208 &0.650 &0.179 &\multicolumn{1}{|c}{0.528}  \\
&CORNIA &0.734 &0.851 &0.567 &0.397 &0.101 &0.831 &0.257 &0.623 &0.218 &0.129 &0.455 &0.123 &\multicolumn{1}{|c}{0.516} \\
&DIIVINE &0.664 &0.738 &0.230 &0.145 &0.062 &0.604 &0.025 &0.391 &0.043 &0.239 &0.661 &0.171 &\multicolumn{1}{|c}{0.479}    \\
&HOSA &0.716 &0.857 &0.721 &0.430 &0.270 &0.775 &0.132 &0.611 &0.189 &0.249 &0.758 &0.201 &\multicolumn{1}{|c}{0.618} \\ 
&SSEQ &0.638 &0.631 &0.365 &0.436 &0.132 &0.651
&-0.115 &0.559 &0.353 &0.247 &0.674 &0.159 &\multicolumn{1}{|c}{0.487} \\ 
&DeepFL-IQA &\bf 0.951 &\bf 0.937 &\bf 0.861 &\bf 0.763 &\bf 0.550 &\bf 0.914 &\bf 0.638 &\bf 0.847 &\bf 0.813 &\bf 0.549 &\bf 0.909 &\bf 0.607   &\multicolumn{1}{|c}{\bf 0.936} \\ \midrule[1.0pt]
\end{tabular}
}
\caption{Performance comparison (SROCC) on all the 25 distortions of the KADID-10k database. Top FR-IQA and NR-IQA models are highlighted.}
\label{tb:kadid10k_comp}
\end{table*}

\begin{table*}[!htbp]
\centering
\resizebox{0.95\textwidth}{!}{
\begin{tabular}{c|l| c c c c c c c c c c c c c} \midrule[1.0pt]
&Method &\bf \#01 &\bf \#02 & \#03 & \#04 & \#05 &\bf \#06 &\bf \#07 &\bf \#08 & \#09 &\bf \#10 &\bf \#11 & \#12& \#13 \\ \midrule[1.0pt] 
\multirow{11}{*}{\rotatebox{90}{FR-IQA}}&SSIM &0.855 &0.752 &0.841 &0.781 &0.865 &0.774 &0.824 &0.969 &0.929 &0.915 &0.950 &0.898 &0.878 \\
& MS-SSIM &0.855 &0.750 &0.848 &0.802 &0.859 &0.783 &0.839 &0.968 &0.931 &0.921 &0.953 &0.899 &0.879\\
& IW-SSIM &0.844 &0.751 &0.817 &0.802 &0.855 &0.728 &0.847 &0.970 &0.915 &0.919 &0.951 &0.839 &0.866 \\ 
&MDSI &\bf 0.948 &\bf 0.879 &\bf 0.946 &0.800 &0.915 &0.860 &0.901 &0.952 &0.950 &0.950 &0.963 &0.890 &0.910 \\ 
&VSI &0.946 &0.871 &0.937 &0.770 &\bf 0.920 &0.874 &0.875 &0.961 &0.948 &0.954 &\bf 0.971 &\bf 0.922 &0.923 \\
&FSIM &0.910 &0.854 &0.890 &0.809 &0.904 &0.825 &0.881 &0.955 &0.933 &0.934 &0.959 &0.861 &0.892 \\
&GMSD &0.946 &0.868 &0.935 &0.708 &0.916 &0.764 &\bf 0.905 &0.911 &\bf 0.952 &0.951 &0.966 &0.840 &0.914\\
&SFF &0.907 &0.817 &0.898 &\bf 0.818 &0.898 &0.787 &0.861 &0.967 &0.909 &0.927 &0.957 &0.883 &0.871 \\
&SCQI &0.945 &0.863 &0.945 &0.817 &0.917 &\bf 0.884 &\bf 0.905 &\bf 0.974 &0.946 &\bf 0.955 &0.966 &0.895 &\bf 0.933\\
&ADD-GSIM &0.936 &0.849 &0.925 &0.750 &0.906 &0.799 &0.899 &0.935 &0.943 &0.944 &0.965 &0.843 &0.916
\\
&SR-SIM &0.925 &0.858 &0.922 &0.786 &0.913 &0.828 &0.849 &0.962 &0.940 &0.939 &0.967 &0.854 &0.917 \\ \hline
\multirow{12}{*}{\rotatebox{90}{NR-IQA}}& BIQI & 0.380 &0.323 &0.533 &0.128 &0.454 &0.533 &0.216
&0.704 &0.506 &0.081 &0.675 &0.307 &0.532 \\
&BLIINDS-II &0.539 &0.560 &0.786 &0.300 &0.726 &0.586 &0.555 &0.830 &0.629 &0.678 &0.854 &0.052 &0.746\\
&BRISQUE &0.429 &0.325 &0.482 &-0.103 &0.684 &0.492 &0.387 &0.710 &0.332 &0.502 &0.671 &-0.042 &0.515 \\
&CORNIA &0.713 &0.100 &0.701 &0.418 &0.879 &0.635 &0.748 &0.835 &0.668 &0.775 &0.757 &0.713
&0.720 \\
&DIIVINE &0.797 &0.456 &0.809 &0.205 &0.867 &0.752 &0.467 &0.864 &0.810 &0.719 &0.813 &0.323 &0.662
\\
&HOSA &0.896 &0.736 &0.842 &0.302 &0.927 &0.606 &0.823 &0.851 &0.823 &0.870 &0.899 & 0.829 &0.568 \\ 
&SSEQ &0.851 &0.585 &0.874 &0.315 &0.861 &0.771 &0.712 &0.713 &0.652 &0.715 &0.817 &-0.004 &0.761 \\ \cline{2-15}
&RankIQA &0.667 &0.620 & 0.821 & 0.365 &0.760 &0.736 &0.783 & 0.809 & 0.767 & 0.866 &0.878 &0.704 &0.810\\ 
&MEON &0.813 &0.722 &0.926 &0.728 &0.911 & 0.901 & 0.888 & \bf0.887 &0.797 &0.850 &0.891 &0.746 &0.716  \\  
&NSSADNN &0.900 &\bf 0.920 &0.920 &\bf 0.860 &0.940 & \bf0.940 &0.880 &0.840 &0.820 &0.860 &0.900 &0.800 &0.720 \\
&Hallucinated-IQA &\bf 0.923 & 0.880 &\bf 0.945 &0.673 &\bf 0.955 &0.810 &0.855 &0.832 &\bf 0.957 &\bf 0.914 &0.624 &0.460 &0.782   \\ 
&DeepFL-IQA &0.906 &0.818 &0.887 &0.499 &0.872 & 0.796 &\bf 0.902 & 0.839 &0.904 &0.855 &\bf 0.913 &\bf 0.868 &\bf 0.820 \\ \midrule[1.0pt]
&Method &\bf \#14 &\bf \#15 &\bf \#16 &\bf \#17 &\bf \#18 &\bf \#19 & \#20 & \#21 & \#22 & \#23 & \#24 &\multicolumn{2}{|c}{Overall} \\ \midrule[1.0pt]
\multirow{11}{*}{\rotatebox{90}{FR-IQA}}&SSIM &0.781 &0.561 &0.759 &0.299 &0.565 &0.765 &0.855 &0.903 &0.830 &0.887 &0.948 &\multicolumn{2}{|c}{0.637} \\ 
& MS-SSIM  &0.794 &0.433 &0.767 &0.466 &0.689 &0.766 &0.857 &0.902 &0.844 &0.892 &0.951 &\multicolumn{2}{|c}{0.786}\\
& IW-SSIM &0.801 &0.372 &0.783 &0.459 &-0.420 &0.773 &0.876 &0.904 &0.840 &0.868 &0.947 &\multicolumn{2}{|c}{0.778}\\
&MDSI &0.822 &\bf 0.693 &0.742 &0.438 &0.800 &0.890 &0.919 &0.956 &0.913 &0.882 &0.964 &\multicolumn{2}{|c}{0.890}    \\ 
&VSI  &0.806 &0.171 &0.770 &0.475 &0.810 &\bf 0.912 &0.924 &0.956
&0.884 &0.891 &0.963 &\multicolumn{2}{|c}{0.897} \\
&FSIM &0.794 &0.553 &0.749 &0.468 &\bf 0.836 &0.857 &0.914 &0.949 &0.882 &0.893 &0.958 &\multicolumn{2}{|c}{0.851}   \\
&GMSD &0.814 &0.663 &0.735 &0.324 &-0.295 &0.889 &\bf 0.930 &0.963 &0.910 &0.853
&\bf 0.968  &\multicolumn{2}{|c}{0.804}  \\
&SFF  &0.767 &0.179 &0.665 &0.469 &0.827 &0.843 &0.901 &0.926 &0.879 &0.879
&0.952 &\multicolumn{2}{|c}{0.851}    \\
&SCQI &0.806 &0.575 &\bf 0.790 &0.448 &0.795 &0.904 &0.918 &0.946 &\bf 0.925 &\bf 0.895 &0.961 &\multicolumn{2}{|c}{\bf 0.905}      \\
&ADD-GSIM &\bf 0.826 &0.609 &0.687 &\bf 0.731 &-0.251 &0.867 &0.925 &\bf 0.967 &0.910 &0.875 &0.963 &\multicolumn{2}{|c}{0.829}   \\
&SR-SIM &0.798 &0.474 &0.661 &0.471 &-0.209 &0.878 &0.926 &0.961 &0.880 &0.875 &0.962 &\multicolumn{2}{|c}{0.807}   \\ \hline
\multirow{12}{*}{\rotatebox{90}{NR-IQA}}& BIQI & 0.158 &0.218 &-0.155 &0.198 &0.160 &0.590 &0.056 &0.580 &0.392 &0.608 &0.835 &\multicolumn{2}{|c}{0.399}  \\
&BLIINDS-II &0.099 &0.417 &0.143 &-0.055 &-0.033 &0.654 &0.123 &0.086 &0.695 &0.468 &0.739 &\multicolumn{2}{|c}{0.557}\\
&BRISQUE &0.175 &0.170 &0.030 &-0.131 &0.294 &0.363 &0.169 &0.059 &0.640 &0.704 &0.666&\multicolumn{2}{|c}{0.415} \\
&CORNIA &0.366 &-0.102 &-0.037 &0.069 &-0.114
&0.702 &0.741 & 0.852 &0.735 &0.611 &0.869 &\multicolumn{2}{|c}{0.612}\\
&DIIVINE &-0.108 &0.334 &0.232 &0.503 &0.083 &0.715 &0.264
&0.753 &0.715 &0.744 &0.859 & \multicolumn{2}{|c}{0.598} \\
&HOSA &0.058 &0.308 &0.247 &0.582 &0.038 &\bf 0.869
&0.645 &0.847 &0.837 &0.789 &\bf 0.902  & \multicolumn{2}{|c}{0.678}         \\ 
&SSEQ &0.062 &0.433 &0.059 &0.289 &-0.111 &0.828
&0.078 &0.591 &0.846 &0.552 &0.831 &   \multicolumn{2}{|c}{0.579}    \\\cline{2-15}
&RankIQA  &0.512 & 0.622 &0.268 &0.613 &0.662 &0.619 &0.644 &0.800 &0.779 &0.629 &0.859 & \multicolumn{2}{|c}{0.780} \\   
&MEON & 0.116 &0.500 &0.177 &0.252 &\bf0.684 &0.849 & 0.406 &0.772 &0.857 &0.779 &0.855 & \multicolumn{2}{|c}{0.808}   \\
&NSSADNN &\bf 0.780 &\bf 0.880 &\bf 0.760 &\bf 0.880 &0.580&\bf0.900&\bf0.880&0.760&\bf0.920&0.740&0.880 &\multicolumn{2}{|c}{0.844} \\
&Hallucinated-IQA  & 0.664 &0.122 &0.182 &0.376 &0.156 &0.850 & 0.614 &0.852 &0.911 &0.381 &0.616 &\multicolumn{2}{|c}{\bf 0.879} \\ 
&DeepFL-IQA & 0.715 & 0.504 & 0.154 &0.787 &0.560 &0.784 &0.824 &\bf 0.907 &0.855 &\bf 0.825 & 0.879 &\multicolumn{2}{|c}{0.858} \\ \midrule[1.0pt]
\end{tabular}
}
\caption{Performance comparison (SROCC) on all the 24 distortions of the TID2013 database. The distortion ID in bold font are included in the KADID-10k database. The performance numbers of deep-learning-based IQA methods are duplicated from their respective papers due to the difficulty for re-implementation. Top FR-IQA and NR-IQA models are highlighted.}
\label{tb:tid2013_comp}
\end{table*}

\newpage
